\documentclass[aps,prc,twocolumn,showpacs,floatfix,nofootinbib,superscriptaddress,amsmath,amssymb,longbibliography,amsproc]{revtex4-1}

\usepackage{bm}
\usepackage{color}
\usepackage{float}
\usepackage{dcolumn}
\usepackage{multirow}
\usepackage[utf8]{inputenc}
\usepackage[english]{babel}
\usepackage[letterpaper, portrait, margin=1in]{geometry}
\usepackage[font={small,it}]{caption}
\usepackage{graphicx}
\usepackage[font={tiny,it}]{subcaption}
\usepackage{nicefrac}
\usepackage{wrapfig}
\usepackage{amsmath}
\usepackage[colorlinks=true,linkcolor=blue,citecolor=blue,urlcolor=blue]{hyperref}

\usepackage{listings}
\usepackage{color}
\definecolor{dkgreen}{rgb}{0,0.6,0}
\definecolor{gray}{rgb}{0.5,0.5,0.5}
\definecolor{mauve}{rgb}{0.58,0,0.82}
\begin{document}

\title{Coupled-Cluster Calculations of Infinite Nuclear Matter in the Complete Basis Limit Using Bayesian Machine Learning}

\thanks{This manuscript has been authored in part by UT-Battelle, LLC, under contract DE-AC05-00OR22725 with the US Department of Energy (DOE). The US government retains and the publisher, by accepting the article for publication, acknowledges that the US government retains a nonexclusive, paid-up, irrevocable, worldwide license to publish or reproduce the published form of this manuscript, or allow others to do so, for US government purposes. DOE will provide public access to these results of federally sponsored research in accordance with the DOE Public Access Plan (http://energy.gov/downloads/doe-public-access-plan).}

\author{Julie Butler}
\affiliation{Department of Biochemistry, Chemistry, and Physics, University of Mount Union, Alliance, Ohio 46601, USA}
\affiliation{Department of Physics and Astronomy and Facility for Rare Isotope Beams, Michigan State University, East Lansing, MI 48824, USA}

\author{Morten Hjorth-Jensen} 
\affiliation{Department of Physics and Center for Computing in Science Education, University of Oslo, N-0316  Oslo, Norway} 
\affiliation{Department of Physics and Astronomy and Facility for Rare Isotope Beams, Michigan State University, East Lansing, MI 48824, USA}

\author{Gustav R. Jansen}
\affiliation{National Center for Computational Sciences, Oak Ridge National Laboratory, Oak Ridge, TN 37831, USA}
\affiliation{Physics Division, Oak Ridge National Laboratory, Oak Ridge, TN 37831, USA}

\begin{abstract}

\textbf{Background} Infinite nuclear matter provides valuable insights into the behavior of nuclear systems and aids our understanding of atomic nuclei and large-scale stellar objects such as neutron stars.  However, partly due to the large basis needed to converge the system's binding energy, size extensive methods such as coupled-cluster theory struggle with long computational run times, even using the nation's largest high-performance computing facilities.

\textbf{Purpose} This research introduces a novel approach to the problem. We propose using a machine learning method to predict the coupled-cluster energies of infinite matter systems in the complete basis limit, leveraging only data collected using smaller basis sets. This method promises to deliver high-accuracy results with significantly reduced run times.

\textbf{Methods} The sequential regression extrapolation (SRE) algorithm, based on Gaussian processes, was created to perform these extrapolations. By combining Bayesian machine learning with a unique method of formatting the training data, we can create a powerful extrapolator that can make accurate predictions given very little data.

\textbf{Results and Discussion} The SRE algorithm successfully predicted the CCD(T) energies for pure neutron matter across six densities near nuclear saturation density, with an average error of 0.0083 MeV/N. The algorithm achieved an average error of 0.038 MeV/A for symmetric nuclear matter. These predictions were made with a time savings of 83.8 node hours for pure neutron matter and 284 node hours for symmetric nuclear matter. Additionally, the symmetry energy at these six densities was predicted with an average error of 0.031 MeV/A and a total time savings of 368 node hours compared to the traditional converged coupled-cluster calculations performed without the SRE algorithm.
\end{abstract}

\maketitle

\section{Introduction\label{sec:intro}}

Nuclear matter is an infinite matter system containing only protons and neutrons, whose interactions are governed by the strong nuclear force and electromagnetic repulsion between the protons \cite{Ref8}. In addition to being an important test bed for nuclear many-body simulations, studies of infinite nuclear matter allow us to gain important insights into nuclear binding and stability, neutron-rich nuclei, nuclear reactions, and neutron stars, among other nuclear processes. Studies of infinite nuclear matter at different proton fractions are crucial in determining the equation-of-state of nuclear matter \cite{Ref3,Ref35,Ref36,Ref37,Ref38,Ref39,Ref41,fore2024}. This work will focus on two cases of infinite nuclear matter: pure neutron matter (with a proton fraction of 0.0) and symmetric nuclear matter (with a proton fraction of 0.5).


Infinite nuclear matter can be studied using different methods. We will use an \textit{ab initio} many-body method called coupled-cluster theory (CC)  in this work \cite{Ref153, Ref152,Ref147, Ref68, Ref16, Ref154, Ref148, Ref149}. Coupled-cluster theory is a powerful many-body method popular in, among others, quantum chemistry, nuclear physics, and condensed matter physics \cite{Ref140, Ref141, Ref149, Ref142,Ref143,Ref145,Ref150,Ref155,Ref7,Ref67,Ref72,Ref74}. It provides a systematic framework for calculating correlations in addition to the mean field, with typical correlations like one-particle one-hole (single excitations or just singles), two-particle-two-hole (doubles), three-particle-three-hole (triples) or more complicated excitations that are summed to infinite order. Coupled-cluster theory, as is also the case with other post Hartree-Fock methods like many-body perturbation theory \cite{Drischler2019,Drischler2021}, Green's function theory \cite{Ref50,Barbieri2017}, in-medium Similarity Renormalization Group theory \cite{Hergert2017} and other approaches \cite{Marino2024}, allows for a systematic inclusion of higher-order correlations not included in a mean-field approach. Here we use the coupled-cluster approximation called coupled-cluster doubles with perturbative triples CCD(T) to calculate the binding energy of infinite nuclear matter \cite{Ref16, Ref158,Ref21,Ref157}\footnote{For infinite matter calculations, one-particle-one-hole excitations (singles) are not present.}. This approximation, which represents state-of-the art infinite matter calculations with CC theory, calculates the correlation energy using two-particle two-hole excitations exactly while including the effects of three-particle three-hole excitations perturbatively. However, though CCD(T) calculations provide the most accurate results, the calculations have a run time that scales as $O(M^6)$ for the iterative part and $O(M^7)$ for the non-iterative part, where $M$ is the number of single-particle states in the calculation. In infinite nuclear matter calculations, the number of single-particle states $M$, which in principle is infinitely large, is typically of the order $M\sim10^3$ or larger in actual implementations. These large single-particle basis sets which in turn define the number of possible many-particle
excitations, are needed in order to obtain converged total energies with respect to the number of single-particle states. Due to the large number of intermediate single-particle states needed for accurate CCD(T) calculations, infinite nuclear matter require many individual calculations. Large-scale studies of infinite nuclear matter, spanning a wide range of densities and proton fractions, have prohibitively expensive computational costs \cite{Ref155}.

Inspired by the recent rise of machine learning, this paper aims to create a machine learning model that can achieve the accuracy of CCD(T) calculations in the complete basis limit but without the prohibitive costs generally associated with these calculations. Machine learning is a fast-evolving field at the intersection of data science and artificial intelligence. It has found many uses in everyday life (chatbots, computer vision, self-driving cars, etc.), but it also has proven itself to be a new and exciting tool in physics. In recent years, physicists have used machine learning to classify images from accelerator experiments, solve differential equations, and accelerate theoretical simulations, among other things, with current funding in almost all disciplines to investigate how to use machine learning for science, see for example Ref.~\cite{Ref210} for a recent review of applications of machine learning methods to nuclear physics and Refs.~\cite{Ref208,Ref209} for applications of machine learning methods to other fields of physics. Specifically, in many-body physics, machine learning has been used to model many-body wavefunctions and extrapolate data to more accurate models \cite{Ref24, Ref25, Ref26, Ref27, Ref28, Ref29, Ref30, Ref31, Ref32,Ref208,Ref211, Ref212, Ref213,Ref214,fore2024}. Inspired by these recent successes in machine learning, we have developed the sequential regression extrapolation (SRE) method which uses the Bayesian machine learning algorithm called Gaussian processes and can accurately predict the binding energies of infinite matter systems in the complete basis limit. Our first application of this method to the homogeneous electron gas in three dimensions \cite{electron_gas_paper} resulted in very promising results for infinite matter studies.

We organize the remainder of this paper into five sections. In section \ref{sec:background}, we introduce the necessary background on infinite nuclear matter and CC theory needed to understand this paper's purpose. In section \ref{sec:methodology}, we develop the SRE method with Gaussian processes for applications to infinite nuclear matter calculations. In section \ref{sec:results}, we show the results of predicting the coupled-cluster energies in the complete basis limit of infinite nuclear matter with the SRE method and discuss the accuracy and time savings that come with these predictions. Finally, section \ref{sec:conclusion} contains our conclusions, investigates the implications of faster infinite nuclear matter calculations in the complete basis limit, and discusses potential future works.

\section{Background\label{sec:background}}

The system of interest in this work, infinite nuclear matter, is an infinite matter system made of nucleons (protons and neutrons) in various fractions which interact via the nuclear force~\cite{Ref8}. Studies of infinite nuclear matter from the many-body perspective are important for understanding the matter within dense astronomical objects such as neutron stars, which offer insights into nuclear processes and astrophysical observables. However, neutrons stars also contain matter that spans several orders of magnitude ($0.1$fm$^{-3}$ and greater) and contain neutrons, protons, electrons, and muons (among other particles) in various fractions. \cite{Ref3,Ref35,Ref36,Ref37,Ref38,Ref39,Ref41,Ref215,Ref216,Ref217}. These particles exist in beta equilibrium ($\beta$-equilibrium) governed by the weak force. For this work, we will restrict our calculations to around nuclear saturation density (0.16 $fm^{-3}$).

Large-scale studies of infinite nuclear matter in various compositions are important for helping to determine the nuclear equation of state (EoS). An accurate determination of the EoS can help studies of neutron stars as it can explain properties such as the relationship between the star's mass and its radius, the thickness of the star's crust, and the rate at which the star will cool down \cite{Ref3}. At the scale of atomic nuclei, determining the EoS also links neutron stars to the neutron skin in atomic nuclei and allow for the calculation of the symmetry energy of nuclear matter. The symmetry energy is crucial because it relates to the difference between proton and neutron radii in nuclei \cite{Ref8}.

For this work, we will limit our calculations to an infinite matter system containing only protons and neutrons in various ratios. The density of protons in the matter, $\rho_p$, compared to the total density, $\rho$, defines the proton fraction, $x_p$. Note that $\rho = \rho_p + \rho_n$, where $\rho_n$ is the density of the neutrons is the matter \cite{Ref3}.

\begin{equation}\label{proton_fraction}
    x_p = \frac{\rho_p}{\rho}.
\end{equation}

 Changing the proton fraction of the infinite nuclear matter system changes its properties. The two types of infinite nuclear matter that we will study in this work are pure neutron matter (PNM), which has $x_p = 0$ and symmetric nuclear matter (SNM), which has $x_p = 0.5$.
 
From the proton fraction, we define here the symmetry energy as the difference between the energy for pure neutron matter and symmetric nuclear matter at a set density \cite{Ref3}:
\begin{equation}
    E_{sym}(\rho) = E(\rho, x_p=1/2) - E(\rho, x_p=0) .
\end{equation}


For an in-depth discussion of the symmetry energy, see, for example, Refs. \cite{symE1} and \cite{symE2}. In this paper, we are interested in calculating the energy of both pure neutron matter and symmetric nuclear matter, as well as the symmetry energy. The nuclear interaction used in all calculations in this paper is NNLO$_{GO}$ \cite{Ref200, Ref201}. This potential is an optimized $\Delta$-full interaction that has been calibrated with nuclear matter properties using 17 low-energy coefficients (LECs) to parameterize the interaction at the next-to-next-to leading order (NNLO) level.

One computational method capable of performing calculations on infinite nuclear matter systems from a many-body approach is coupled-cluster theory (CC) \cite{Ref153, Ref152, Ref147, Ref68, Ref16,
Ref154, Ref148, Ref149}. Coupled-cluster theory begins with the exponential ansatz, which states that any many-body wavefunction, $|\Psi\rangle$, can be written as the Fermi vacuum state, $|\Phi_0\rangle$, acted on by the exponential operator $e^{\hat{T}}$:

\begin{equation}\label{eq:exponential}
	|\Psi\rangle = e^{\hat{T}}|\Phi_0\rangle.
\end{equation}

The operator $\hat{T}$ is known as the cluster or correlation operator, and is the sum of the $N$ i-particle i-hole excitation operators, where N is the number of particles in the system:

\begin{equation} \label{eq:cluster}
	\hat{T} = \sum_{i=1}^{N}\hat{T}_i.
\end{equation}

If Eq.~(\ref{eq:cluster}) is used in its entirety in a coupled-cluster calculation, then the result of the calculation is equivalent to the result from solving the Schr\"{o}dinger equation for the system. However, the cluster operator can rarely be entirely used due to the immense computational costs of higher-order excitation operators. Fortunately, coupled-cluster theory has a simple and physically motivated truncation scheme in which excitation operators over a set level are forced to zero. Thus, we develop the coupled-cluster singles method when $\hat{T}\approx\hat{T}_1$, the coupled-cluster singles and doubles method when $\hat{T}\approx\hat{T}_1+\hat{T}_2$, and the coupled-cluster singles, doubles, and triples method when
$\hat{T}\approx\hat{T}_1+\hat{T}_2+\hat{T}_3$ \cite{Ref155, Ref157}. With the inclusion of triples correlations, coupled-cluster theory allows for a computationally efficient inclusion of important many-body correlations beyond a mean-field approach.


Even though we are required to truncate the cluster operator due to
computational limitations, these truncated cluster operators can still have very long run times. A CCSD calculation has an expected run time of $O(M^6)$, where M is the number of single-particle states in the calculation, but a CCSDT calculation has an expected run time of $O(M^8)$. In nuclear many-body calculations, with complex interactions and large single-particle bases, CCSDT calculations are rarely computationally feasible. However, since many aspects of the nuclear interaction emerge at the three-body order or higher, we still want to include aspects of the $\hat{T}_3$ operator in our calculations, but with run times more similar to CCSD calculations. One method that can be used to bridge the gap between CCSD and CCSDT is called coupled-cluster singles and doubles with perturbative triples, CCSD(T). In a CCSD(T) calculation, some aspects of the $\hat{T}_3$ operator are approximated with the $\hat{T}_2$ operator, but the run time is an iterative $O(M^6)$ (computing the $\hat{T_1}$ and $\hat{T}_2$ operators) and a non-iterative component with a run time of $O(M^7)$ (approximating $\hat{T}_3$). Although these run times are higher than for a CCSD calculation, they are significantly lower than the run times for CCSDT calculations, and thus CCSD(T) is the most accurate and accessible coupled-cluster method commonly used in nuclear applications. Note that, for infinite matter systems, the $\hat{T}_1$ operator has no contribution due to symmetry in the momentum of the system, so the CCSD(T) method simply becomes CCD(T).

\begin{figure*}[htb]
\centering
\begin{subfigure}[t]{0.5\textwidth}
  \centering
  \includegraphics[width=\linewidth]{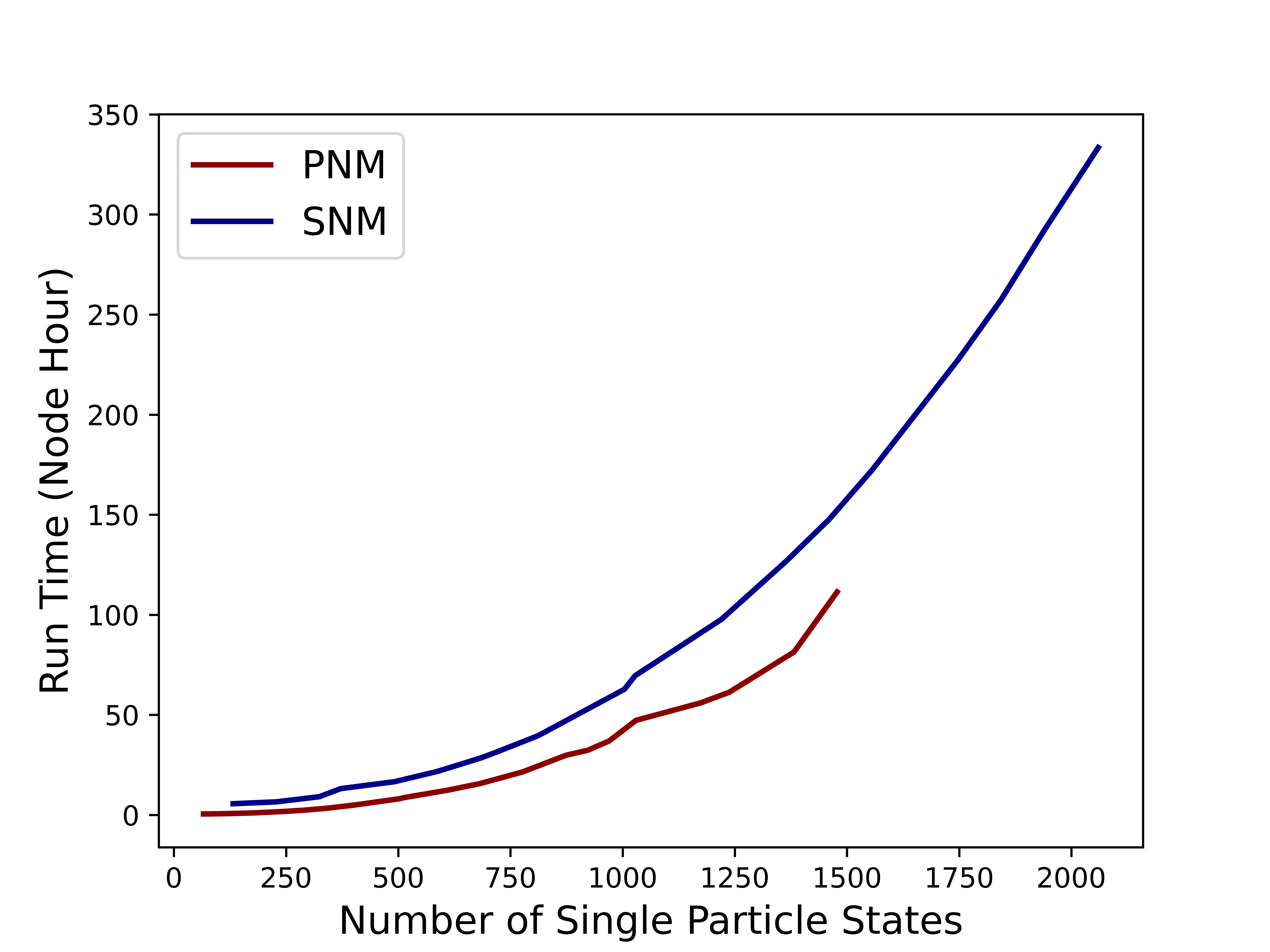}
  \label{fig:times}
\end{subfigure}%
\begin{subfigure}[t]{0.5\textwidth}
  \centering
  \includegraphics[width=\linewidth]{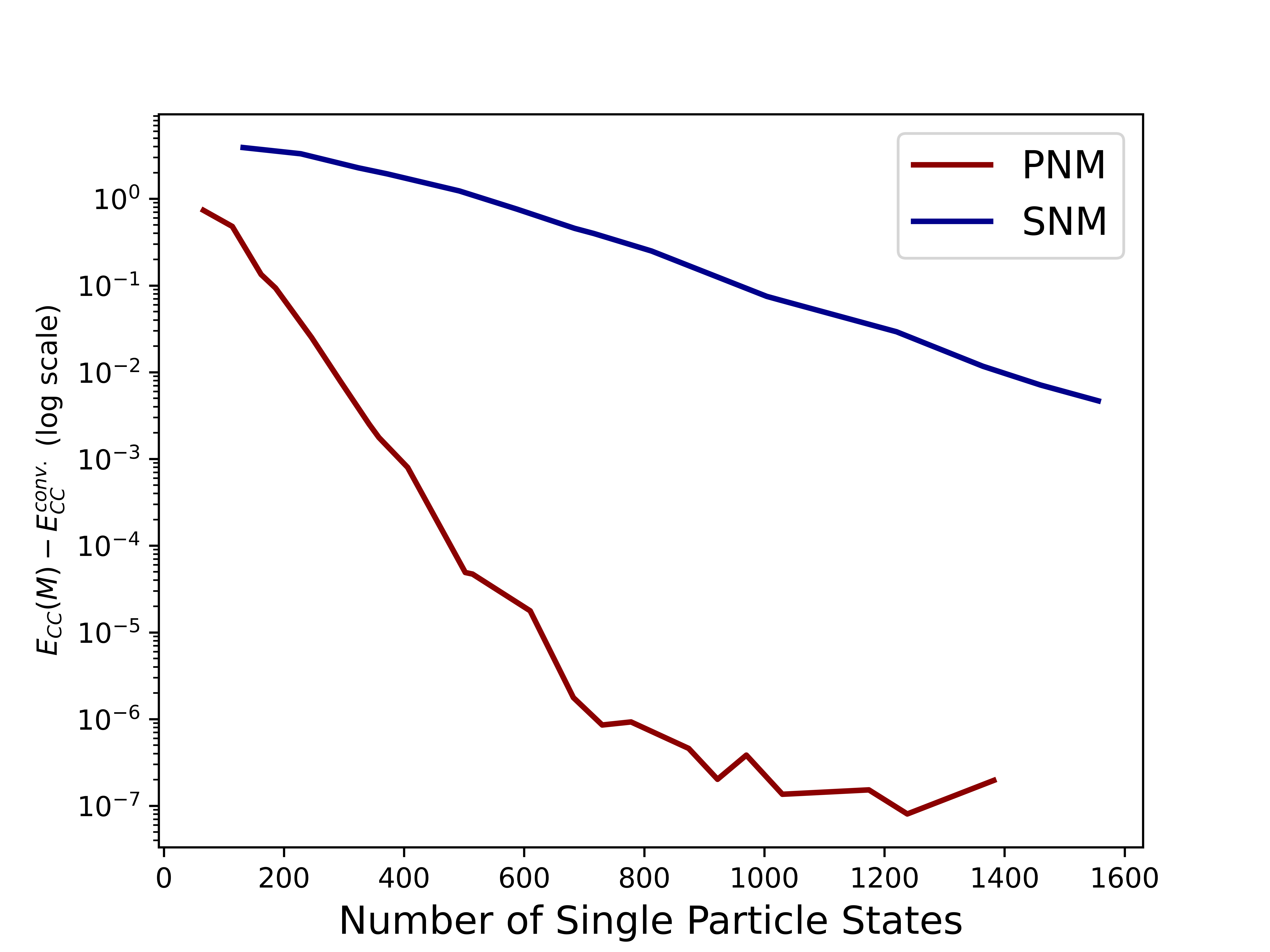}
  \label{fig:accuracy}
\end{subfigure}
\caption{As the number of single-particle states in an infinite nuclear matter calculation increases, the run time of the calculation drastically increases (left), but the error of the calculation (compared to the converged results) decreases (right).}
\label{fig:times_and_accuracy}
\end{figure*}

Even with the lessened run time of a CCD(T) calculation, it is still computationally difficult to perform calculations at a high enough number of single-particle states that the energy of the system converges (the complete basis limit). Figure \ref{fig:times_and_accuracy} shows the run time for CCD(T) calculations as a function of the number of single-particle
states for both pure neutron matter and symmetric nuclear matter at N = 66 neutrons (132 nucleons) and $\rho = 0.16fm^{-3}$. These
calculations were run with an Apple Mac Studio using 3.5 GHz M2 Ultra processor. The coupled-cluster code was parallelized and each
calculation was performed with 20 compute nodes. We report run times in units of node hours, defined as the run time of the calculation (in hours) multiplied by the number of nodes used in the calculation. This was done to better reflect the true computational time needed by these calculations. As shown in Fig.~\ref{fig:times_and_accuracy}, as the number of single-particle states in the calculation increases, the run time of the calculation also drastically increases. However, as we can see in Fig.~\ref{fig:times_and_accuracy}, as the number of single-particle states in the calculations is increased, the accuracy of the calculation (compared to results in the complete basis limit) also increases. Therefore, if we want high accuracy in our calculations we need to perform calculations at a high number of single-particle states.

While we want to perform the CCD(T) calculations at a high number of single-particle states ($M = 1,478$ for pure neutron matter and $M = 2,060$ for symmetric nuclear matter), the high computational costs at these values of $M$ make calculations difficult if not computationally prohibitive (14 node hours for pure neutron matter and 65 hours for symmetric nuclear matter per calculation). These high computational costs are especially problematic considering that studies of infinite nuclear matter often require calculations at many different densities and proton fractions, making hundreds of node hours per calculation an unattractive prospect. Therefore, in this work, we want to develop a machine learning method which is capable of predicting the energy of these infinite nuclear matter systems at high numbers of single particle states, but we only want to train the model on calculations at low numbers of single particle states. However, many existing machine learning methods require a large amount of training data, which would be very difficult to calculate in this case. Therefore, we will use the sequential regression extrapolation (SRE) method, developed in Ref.~\cite{my_thesis} and Ref.~\cite{electron_gas_paper}, to perform extrapolations with only a very small training set.

\section{Methodology\label{sec:methodology}}

As is shown in Fig.~\ref{fig:times_and_accuracy}, as the number of
single particle states in the calculation increases, so does the
accuracy of the calculation.  Phrased another way, as the number of
single particle states in the calculation increases, the CCD(T) energy (or correlation energy) converges to a constant value.  This is also a property of another many-body method called many-body perturbation theory (MBPT).  When performing a CCD(T) calculation, all of the diagrams needed to find the MBPT energy to the second order (MBPT2) are computed as well, thus the MBPT2 energy (or correlation energy) can be found from a CCD(T) calculation with no additional computational time. We can use the MBPT2 correlation energy as an estimate of the system's correlation energy. Even though MBPT2 calculations of infinite nuclear matter can be computationally expensive, they are not nearly as expensive as CCD(T) calculations of the same system and provide a reliable estimate of the intermediate states. Thus, we aim to develop a method which can predict the CCD(T) energy of a system in the complete basis limit, using only MBPT2 calculations and CCD(T) calculations at smaller basis sizes. 


Since we know that both of these correlation energies (CCD(T) and MBPT2) converge at some large number of single particle states, then at that value of $M$, the ratio of the two correlation energies is a constant which we will refer to as $m$.

\begin{equation}\label{eq:ratio}
    \lim_{M\rightarrow\infty} \frac{\Delta E_{CC,M}}{\Delta E_{MBPT2,M}} = m.
\end{equation}

The constant $m$ is the value we want to predict with our SRE
algorithm, since given this value and the MBPT2 calculation at a high value of $M$, we can predict the CCD(T) correlation energy in the complete basis limit. Therefore, the training data we need to generate for our model are calculations for the CCD(T) and MBPT2 correlation energies at a set density and number of particles, but at increasing numbers of single particle states. For the pure neutron matter results, we used three training points per density, with the number of single particle states ranging from 186 to 294.  For the symmetric nuclear matter calculations, we used six training points per calculations, with the number of single particle states ranging from 324 to 684.  The training data consisted of a CCD(T) calculation and a MBPT2 calculation at the same number of single particle states, divided by each other to make the ratio in Eq.~(\ref{eq:ratio}) and arranged in sequential and ascending order.

\begin{equation}
    y = \frac{\Delta E_{CC,M_k}}{\Delta E_{CC,M_k}} = \frac{\Delta E_{CC,M_1}}{\Delta E_{CC,M_1}}, \frac{\Delta E_{CC,M_2}}{\Delta E_{CC,M_2}}, \frac{\Delta E_{CC,M_3}}{\Delta E_{CC,M_3}} ...
\end{equation}

First, we train our Gaussian process (GP) \cite{Ref218,Ref219} algorithm by feeding it one ratio as the input and having it learn how to predict the next ratio in the sequence as an output. The kernel used by the GP algorithms in this study is a rational quadratic kernel, which has been modified by a white kernel and a constant kernel

\begin{equation}
	f_{GP}(\frac{\Delta E_{CC, k-1}}{\Delta E_{MBPT, k-1}}) = \frac{\Delta E_{CC, k}}{\Delta E_{MBPT, k}}.
\end{equation}

We then take the trained model and use it to generate the ratios of
energies a large number of times until the ratio has converged.  This is the same constant from Eq.~\eqref{eq:ratio}

\begin{equation}
	\lim_{k\rightarrow\infty} \frac{\Delta E_{CC, k}}{\Delta E_{MBPT, k}} = m.
\end{equation}

The constant $m$ can then be multiplied by the MBPT2 correlation energy at a high number of single particle states to approximate the coupled-cluster correlation energy at that same high number of single particle states

\begin{equation}
	m\Delta E_{MBPT, Large\ M} = \Delta E_{CC, Large\ M}.
\end{equation}

Once the correlation energy has been approximated, it can be added
back to the reference energy, found through a Hartree-Fock
calculation, to determine the total coupled-cluster energy
\begin{equation}
	E_{CC} = E_0 + \Delta E_{CC}.
\end{equation}

Performing the extrapolations with the correlation energies instead of the full energies leads to more accurate predictions \cite{my_thesis}. Calculations for this work were performed using 66 neutrons for PNM and 132 nucleons for SNM. At these particle numbers the correlation energies of the system are very close to what they are at the thermodynamic limit \cite{Ref8}.  As a final note, for all densities considered in this work, there are no problems with achieving converged results for MBPT to second order.  In contrast, in our work on the electron gas \cite{electron_gas_paper}, the range of studied densities was limited to high densities due to converge problems of MBPT2 as function of the number of single-particle basis states $M$.

\section{Results and Discussion\label{sec:results}}
\begin{figure*}[htb]
\centering
\begin{subfigure}{.5\textwidth}
  \centering
  \includegraphics[width=\linewidth]{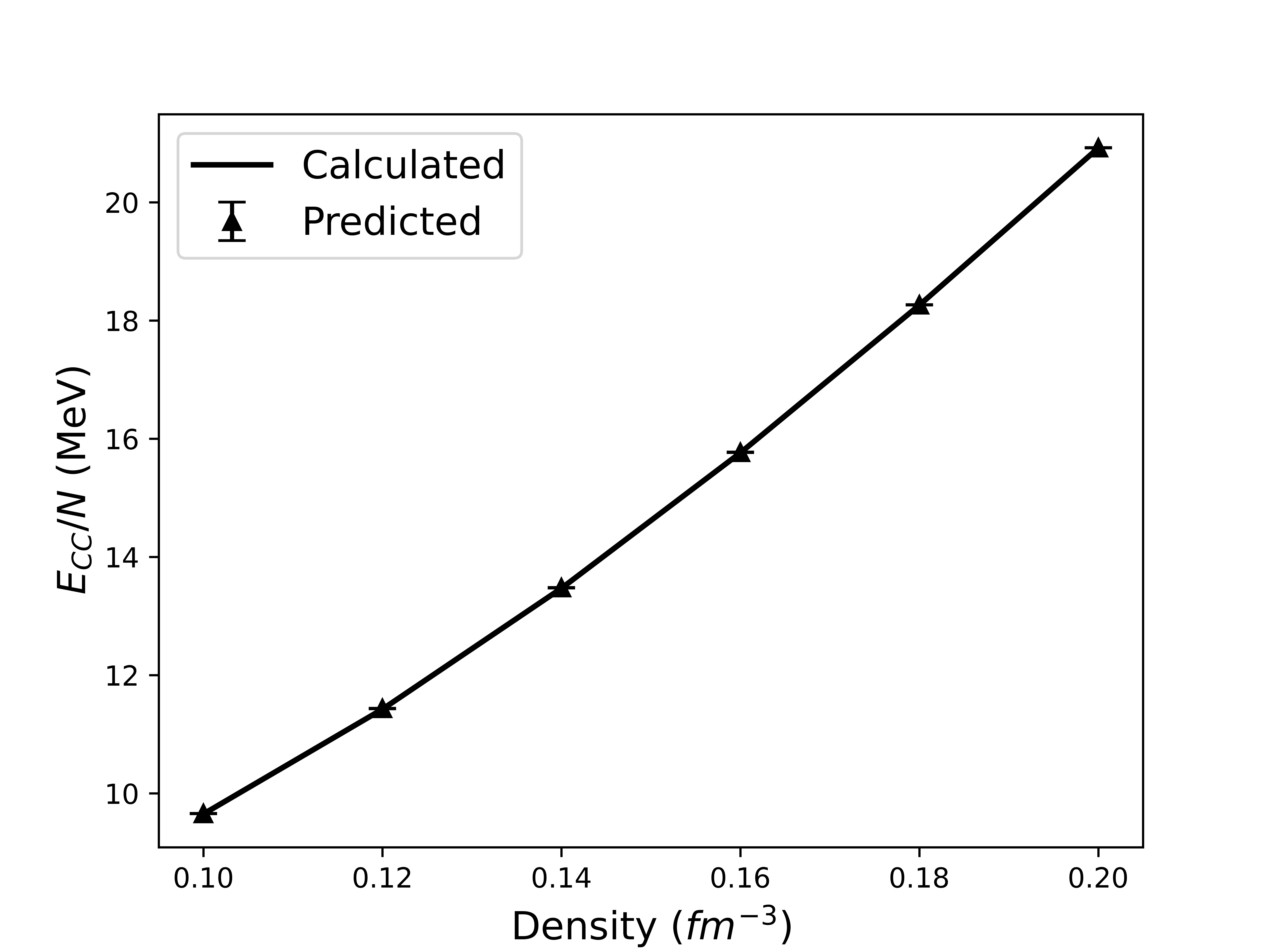}
  \label{fig:energies_pnm}
\end{subfigure}%
\begin{subfigure}{.5\textwidth}
  \centering
  \includegraphics[width=\linewidth]{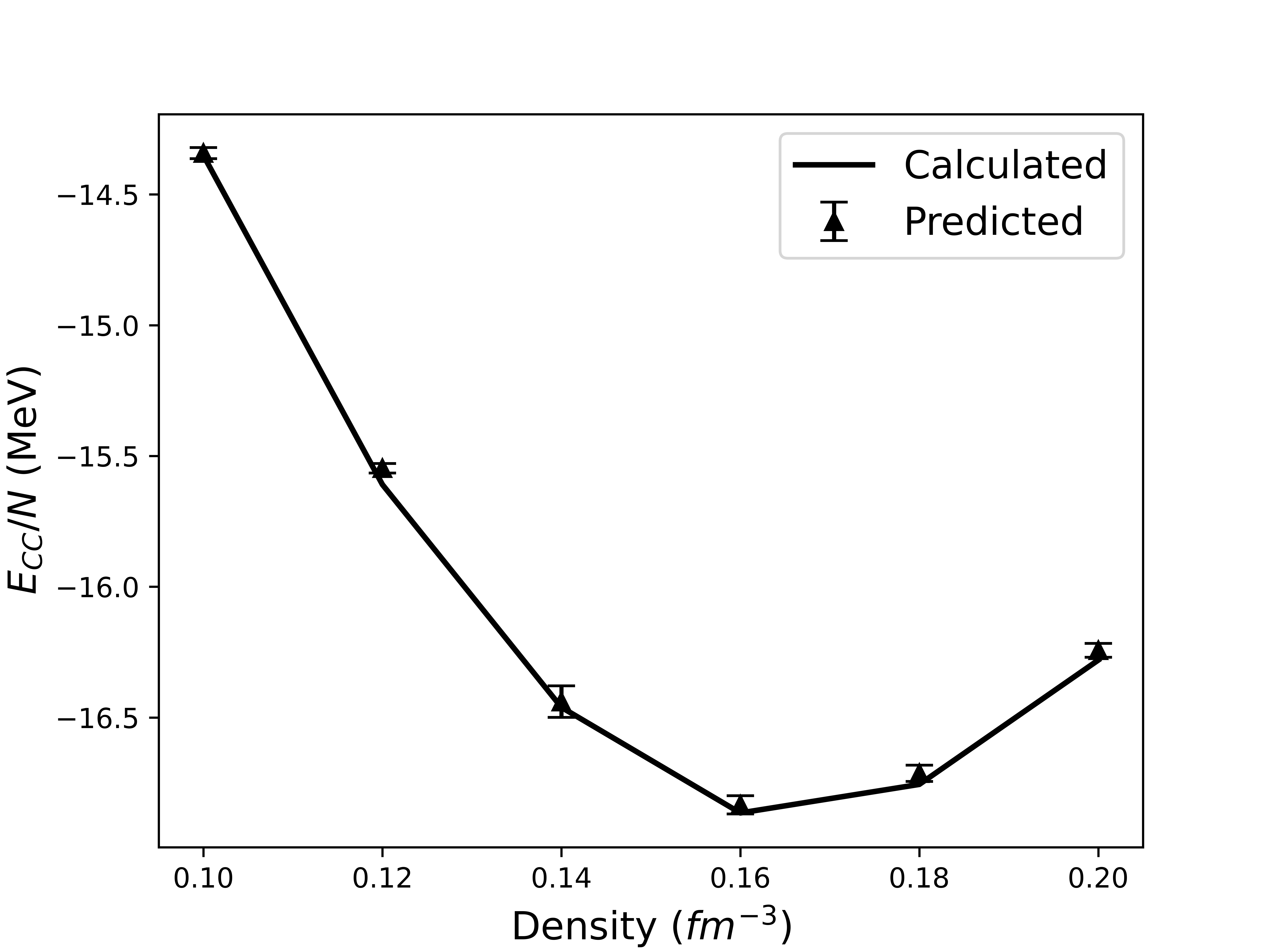}
  \label{fig:energies_snm}
\end{subfigure}
\caption{A comparison between the CCD(T) energies calculated at convergence (solid line) and the energies predicted by the SRE method (triangular markers) for pure neutron matter (left) and symmetric nuclear matter (right) for six densities around nuclear density. The average error between the predicted and calculated data for pure neutron matter is 0.0083 MeV/N and is 0.038 MeV/A for symmetric nuclear matter.}
\label{fig:energies}
\end{figure*}

To ensure that the SRE method is able to improve the accuracy of the CCD(T) energy results compared to its training data, the average percent error between the converged energies and the energies calculated at $M = 294$ for pure neutron matter is 21.4$\%$. For symmetric nuclear matter, the average percent error between the converged energies and the results at $M = 358$ is 9.34$\%$.  These are the most accurate results we can achieve with the training data; if the SRE method is able to achieve more accurate results than these percent errors, then we have improved the accuracy with limited additional computational costs. In Fig.~\ref{fig:energies}, we plot the pure neutron matter results on the left and the symmetric nuclear matter matter results on the right, with the solid line representing the converged energies that have been fully calculated and the triangular markers representing the SRE predictions. The error bars on the SRE predictions represent the uncertainties on the prediction that comes from the Gaussian process algorithm.

In Fig.~\ref{fig:energies}, the average error for the pure neutron
matter calculations is $0.0083$ MeV$/N$, where $N$ is the number of
neutrons, and the average error for the symmetric nuclear matter
calculations is $0.038$ MeV$/A$ where $A$ is the number of
nucleons. This corresponds to an average percent error between the
predicted and the calculated results of 0.057 $\%$ for pure neutron
matter and 0.21$\%$ for symmetric nuclear matter. Additionally, the
average value for the uncertainty on the predictions for pure neutron matter is 0.0043 MeV$/N$ and is 0.032 MeV$/A$ for symmetric nuclear matter.  Both of these uncertainties are a very small percentage of the overall coupled-cluster energy. Based on these quantitative results and the close matches of the plots in
Fig.~\ref{fig:energies}, it appears that the SRE method can very
accurately predict the converged CCD(T) energies for infinite nuclear matter systems, given only the unconverged results.

    When considering the time savings of predicting the CCD(T)
energies over calculating them at convergence, it requires 84.3 node hours to calculate the six fully converged data points in
Fig.~\ref{fig:energies} for pure neutron matter. The time needed to
generate the training data for the SRE method to predict all six data points is 0.41 node hours, with the machine learning process adding a negligible amount of time.  Thus the time saved with predicting the CCD(T) energies for pure neutron matter with the SRE method is 83.8 node hours for six data points, or 3.49 \textit{node days} of computational time saved. For symmetric nuclear matter, it requires 390 node hours to calculate the six data points at convergence and 106 node hours to generate the training data for the SRE predictions (again the machine learning process adds a negligible amount of time). This leads to a total time savings of 284 node hours in generating the six symmetric nuclear data points, or, on average, a time savings of \textit{11.8} node days. This large time savings with just six calculations per proton fraction offers great promise to make large sale studies of infinite nuclear matter, such as those needed to study the nuclear equation of state, computationally feasible.

\subsection{Symmetry Energy}

From the results shown in Fig.~\ref{fig:energies} we are able to
calculate the symmetry energy for infinite nuclear matter. The
symmetry energy is defined as the difference between the energy for
pure neutron matter and the energy for symmetric nuclear matter.


\begin{figure}[htb]
    \centering
    \includegraphics[width=\linewidth]{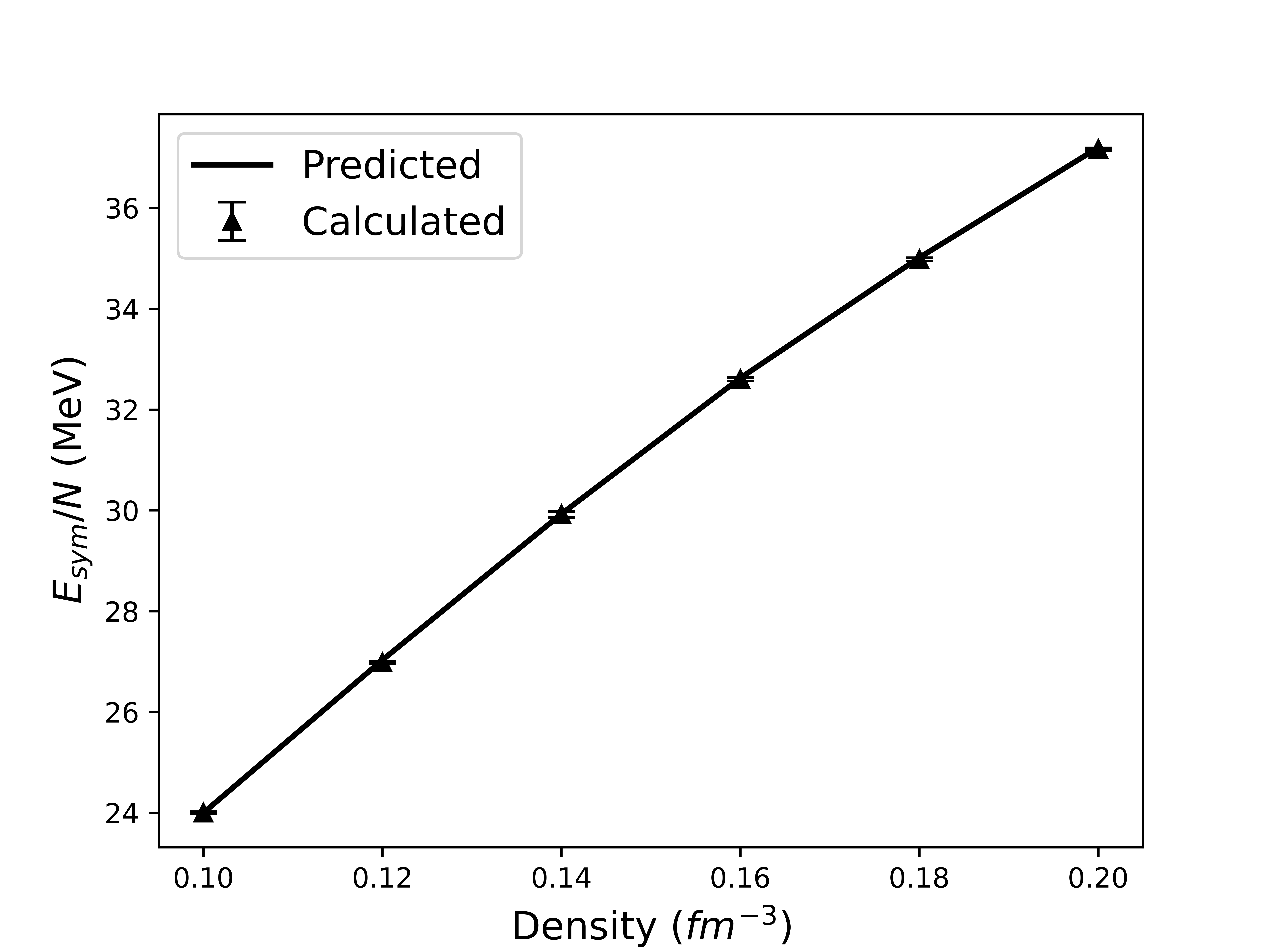}
    \caption{The symmetry energy calculated at six different densities around nuclear density. The solid line depicts the results from computing the symmetry energy with energies calculated at convergence and the triangular markers show the symmetry energy calculated with SRE predictions. The error bars on the triangular markers are the uncertainties on the predictions from the Gaussian process.}
    \label{fig:symmetry}
\end{figure}

For the six symmetry energy calculations in Fig.~\ref{fig:symmetry}, the average error is 0.031 MeV$/A$, which equates to an average percent error of 0.084$\%$.  The average value for the uncertainties on the predictions is 0.038 MeV$/A$.  Additionally, the time saved when performing the SRE predictions over performing the CCD(T) calculations to convergence is 368 node hours, or over 15.3 node days of computational time saved.

\section{Conclusion\label{sec:conclusion}}

Large-scale studies of infinite nuclear matter systems have the
ability to advance many areas of study within the field of nuclear
physics, but the high computational costs associated with these
calculations are a bottleneck to many of these studies. The SRE method developed in this paper has the ability to make these studies more feasible. Though this paper only considers calculations at six different densities, the computational time savings is many weeks. This type of time savings has the ability to make previously impossible large-scale studies of infinite nuclear matter much more computationally feasible.

While we present the SRE method in this paper in the light of
coupled-cluster calculations of infinite nuclear matter, the SRE
method is extremely general and has many further applications. It has already been shown that the method can be used for basis completeness extrapolations on coupled-cluster calculations for other systems (see Ref. \cite{electron_gas_paper} for an application to the homogeneous electron gas). While both this paper and Ref. \cite{electron_gas_paper} apply the SRE method to infinite matter systems, the method can also be applied to calculations of finite systems, such as, large and deformed atomic nuclei, which also suffer from basis truncation errors. Additionally, the method does not need to be restrained to coupled-cluster calculations and can be modified to work with other many-body methods as well, such as many-body perturbation theory and the in-medium similarity renormalization group theory \cite{Ref48, Ref65}. Therefore, the SRE approach, capable of performing precise extrapolations with limited data sets, holds the potential to reduce computational runtime and resource demands in calculations across the realm of many-body physics, rendering large-scale and complex system studies more achievable.

All code and data needed to recreate the plots and calculations in this paper are available at www.github.com/butler-julie/ INM-SRE.

\acknowledgements

This research is supported by the U.S. National Science Foundation
Grants No. PHY-2310020 and PHY-2013047. This material is based upon
work supported by the U.S. Department of Energy, Office of Science,
Office of Workforce Development for Teachers and Scientists, Office of Science Graduate Student Research (SCGSR) program. The SCGSR program is administered by the Oak Ridge Institute for Science and Education (ORISE) for the DOE. ORISE is managed by ORAU under contract number DE-SC0014664. All opinions expressed in this paper are the author’s and do not necessarily reflect the policies and views of DOE, ORAU, or ORISE. This research used resources of the Oak Ridge Leadership Computing Facility located at Oak Ridge National Laboratory, which is supported by the Office of Science of the Department of Energy under contract No. DE-AC05-00OR22725. Computer time was provided by the Innovative and Novel Computational Impact on Theory and Experiment (INCITE) program.

    \bibliography{main}

\begin{thebibliography}{66}%
\makeatletter
\providecommand \@ifxundefined [1]{%
 \@ifx{#1\undefined}
}%
\providecommand \@ifnum [1]{%
 \ifnum #1\expandafter \@firstoftwo
 \else \expandafter \@secondoftwo
 \fi
}%
\providecommand \@ifx [1]{%
 \ifx #1\expandafter \@firstoftwo
 \else \expandafter \@secondoftwo
 \fi
}%
\providecommand \natexlab [1]{#1}%
\providecommand \enquote  [1]{``#1''}%
\providecommand \bibnamefont  [1]{#1}%
\providecommand \bibfnamefont [1]{#1}%
\providecommand \citenamefont [1]{#1}%
\providecommand \href@noop [0]{\@secondoftwo}%
\providecommand \href [0]{\begingroup \@sanitize@url \@href}%
\providecommand \@href[1]{\@@startlink{#1}\@@href}%
\providecommand \@@href[1]{\endgroup#1\@@endlink}%
\providecommand \@sanitize@url [0]{\catcode `\\12\catcode `\$12\catcode `\&12\catcode `\#12\catcode `\^12\catcode `\_12\catcode `\%12\relax}%
\providecommand \@@startlink[1]{}%
\providecommand \@@endlink[0]{}%
\providecommand \url  [0]{\begingroup\@sanitize@url \@url }%
\providecommand \@url [1]{\endgroup\@href {#1}{\urlprefix }}%
\providecommand \urlprefix  [0]{URL }%
\providecommand \Eprint [0]{\href }%
\providecommand \doibase [0]{http://dx.doi.org/}%
\providecommand \selectlanguage [0]{\@gobble}%
\providecommand \bibinfo  [0]{\@secondoftwo}%
\providecommand \bibfield  [0]{\@secondoftwo}%
\providecommand \translation [1]{[#1]}%
\providecommand \BibitemOpen [0]{}%
\providecommand \bibitemStop [0]{}%
\providecommand \bibitemNoStop [0]{.\EOS\space}%
\providecommand \EOS [0]{\spacefactor3000\relax}%
\providecommand \BibitemShut  [1]{\csname bibitem#1\endcsname}%
\let\auto@bib@innerbib\@empty
\bibitem [{\citenamefont {Baardsen}\ \emph {et~al.}(2013)\citenamefont {Baardsen}, \citenamefont {Ekstr\"om}, \citenamefont {Hagen},\ and\ \citenamefont {Hjorth-Jensen}}]{Ref8}%
  \BibitemOpen
  \bibfield  {author} {\bibinfo {author} {\bibfnamefont {G.}~\bibnamefont {Baardsen}}, \bibinfo {author} {\bibfnamefont {A.}~\bibnamefont {Ekstr\"om}}, \bibinfo {author} {\bibfnamefont {G.}~\bibnamefont {Hagen}}, \ and\ \bibinfo {author} {\bibfnamefont {M.}~\bibnamefont {Hjorth-Jensen}},\ }\href {\doibase 10.1103/PhysRevC.88.054312} {\bibfield  {journal} {\bibinfo  {journal} {Phys. Rev. C}\ }\textbf {\bibinfo {volume} {88}},\ \bibinfo {pages} {054312} (\bibinfo {year} {2013})}\BibitemShut {NoStop}%
\bibitem [{\citenamefont {Lietz}\ \emph {et~al.}(2017)\citenamefont {Lietz}, \citenamefont {Novario}, \citenamefont {Jansen}, \citenamefont {Hagen},\ and\ \citenamefont {Hjorth-Jensen}}]{Ref3}%
  \BibitemOpen
  \bibfield  {author} {\bibinfo {author} {\bibfnamefont {J.~G.}\ \bibnamefont {Lietz}}, \bibinfo {author} {\bibfnamefont {S.}~\bibnamefont {Novario}}, \bibinfo {author} {\bibfnamefont {G.~R.}\ \bibnamefont {Jansen}}, \bibinfo {author} {\bibfnamefont {G.}~\bibnamefont {Hagen}}, \ and\ \bibinfo {author} {\bibfnamefont {M.}~\bibnamefont {Hjorth-Jensen}},\ }\enquote {\bibinfo {title} {Computational nucl. phys. and post hartree-fock methods},}\ in\ \href {\doibase 10.1007/978-3-319-53336-0_8} {\emph {\bibinfo {booktitle} {An Advanced Course in Computational Nucl. Phys. : Bridging the Scales from Quarks to Neutron Stars}}},\ \bibinfo {editor} {edited by\ \bibinfo {editor} {\bibfnamefont {M.}~\bibnamefont {Hjorth-Jensen}}, \bibinfo {editor} {\bibfnamefont {M.~P.}\ \bibnamefont {Lombardo}}, \ and\ \bibinfo {editor} {\bibfnamefont {U.}~\bibnamefont {van Kolck}}}\ (\bibinfo  {publisher} {Springer International Publishing},\ \bibinfo {address} {Cham},\ \bibinfo {year} {2017})\ pp.\ \bibinfo {pages}
  {293--399}\BibitemShut {NoStop}%
\bibitem [{Ref(1983)}]{Ref35}%
  \BibitemOpen
  \enquote {\bibinfo {title} {Neutron star models: Masses and radii},}\ in\ \href {\doibase 10.1002/9783527617661.ch9} {\emph {\bibinfo {booktitle} {Black Holes, White Dwarfs, and Neutron Stars}}}\ (\bibinfo  {publisher} {John Wiley \& Sons, Ltd},\ \bibinfo {year} {1983})\ Chap.~\bibinfo {chapter} {9}, pp.\ \bibinfo {pages} {241--266}\BibitemShut {NoStop}%
\bibitem [{\citenamefont {Lattimer}\ and\ \citenamefont {Prakash}(2001)}]{Ref36}%
  \BibitemOpen
  \bibfield  {author} {\bibinfo {author} {\bibfnamefont {J.~M.}\ \bibnamefont {Lattimer}}\ and\ \bibinfo {author} {\bibfnamefont {M.}~\bibnamefont {Prakash}},\ }\href {\doibase 10.1086/319702} {\bibfield  {journal} {\bibinfo  {journal} {Ap. J.}\ }\textbf {\bibinfo {volume} {550}},\ \bibinfo {pages} {426} (\bibinfo {year} {2001})}\BibitemShut {NoStop}%
\bibitem [{\citenamefont {Lattimer}\ and\ \citenamefont {Prakash}(2007)}]{Ref37}%
  \BibitemOpen
  \bibfield  {author} {\bibinfo {author} {\bibfnamefont {J.~M.}\ \bibnamefont {Lattimer}}\ and\ \bibinfo {author} {\bibfnamefont {M.}~\bibnamefont {Prakash}},\ }\href {\doibase 10.1016/j.physrep.2007.02.003} {\bibfield  {journal} {\bibinfo  {journal} {Phys. Rep.}\ }\textbf {\bibinfo {volume} {442}},\ \bibinfo {pages} {109} (\bibinfo {year} {2007})},\ \bibinfo {note} {the Hans Bethe Centennial Volume 1906-2006}\BibitemShut {NoStop}%
\bibitem [{\citenamefont {Steiner}\ \emph {et~al.}(2010)\citenamefont {Steiner}, \citenamefont {Lattimer},\ and\ \citenamefont {Brown}}]{Ref38}%
  \BibitemOpen
  \bibfield  {author} {\bibinfo {author} {\bibfnamefont {A.~W.}\ \bibnamefont {Steiner}}, \bibinfo {author} {\bibfnamefont {J.~M.}\ \bibnamefont {Lattimer}}, \ and\ \bibinfo {author} {\bibfnamefont {E.~F.}\ \bibnamefont {Brown}},\ }\href {\doibase 10.1088/0004-637X/722/1/33} {\bibfield  {journal} {\bibinfo  {journal} {Ap. J.}\ }\textbf {\bibinfo {volume} {722}},\ \bibinfo {pages} {33} (\bibinfo {year} {2010})}\BibitemShut {NoStop}%
\bibitem [{\citenamefont {Lattimer}(2012)}]{Ref39}%
  \BibitemOpen
  \bibfield  {author} {\bibinfo {author} {\bibfnamefont {J.~M.}\ \bibnamefont {Lattimer}},\ }\href {\doibase 10.1146/annurev-nucl-102711-095018} {\bibfield  {journal} {\bibinfo  {journal} {Ann. Rev. Nucl. Part. Sci.}\ }\textbf {\bibinfo {volume} {62}},\ \bibinfo {pages} {485} (\bibinfo {year} {2012})}\BibitemShut {NoStop}%
\bibitem [{\citenamefont {Heiselberg}\ and\ \citenamefont {Hjorth-Jensen}(2000)}]{Ref41}%
  \BibitemOpen
  \bibfield  {author} {\bibinfo {author} {\bibfnamefont {H.}~\bibnamefont {Heiselberg}}\ and\ \bibinfo {author} {\bibfnamefont {M.}~\bibnamefont {Hjorth-Jensen}},\ }\href {\doibase 10.1016/S0370-1573(99)00110-6} {\bibfield  {journal} {\bibinfo  {journal} {Phys. Rep.}\ }\textbf {\bibinfo {volume} {328}},\ \bibinfo {pages} {237} (\bibinfo {year} {2000})}\BibitemShut {NoStop}%
\bibitem [{\citenamefont {Fore}\ \emph {et~al.}(2024)\citenamefont {Fore}, \citenamefont {Kim}, \citenamefont {Hjorth-Jensen},\ and\ \citenamefont {Lovato}}]{fore2024}%
  \BibitemOpen
  \bibfield  {author} {\bibinfo {author} {\bibfnamefont {B.}~\bibnamefont {Fore}}, \bibinfo {author} {\bibfnamefont {J.}~\bibnamefont {Kim}}, \bibinfo {author} {\bibfnamefont {M.}~\bibnamefont {Hjorth-Jensen}}, \ and\ \bibinfo {author} {\bibfnamefont {A.}~\bibnamefont {Lovato}},\ }\href {\doibase 10.48550/arXiv.2407.21207} {\enquote {\bibinfo {title} {Investigating the crust of neutron stars with neural-network quantum states},}\ } (\bibinfo {year} {2024})\BibitemShut {NoStop}%
\bibitem [{\citenamefont {Coester}(1958)}]{Ref153}%
  \BibitemOpen
  \bibfield  {author} {\bibinfo {author} {\bibfnamefont {F.}~\bibnamefont {Coester}},\ }\href {\doibase 10.1016/0029-5582(58)90280-3} {\bibfield  {journal} {\bibinfo  {journal} {Nucl. Phys.}\ }\textbf {\bibinfo {volume} {7}},\ \bibinfo {pages} {421} (\bibinfo {year} {1958})}\BibitemShut {NoStop}%
\bibitem [{\citenamefont {Coester}\ and\ \citenamefont {Kümmel}(1960)}]{Ref152}%
  \BibitemOpen
  \bibfield  {author} {\bibinfo {author} {\bibfnamefont {F.}~\bibnamefont {Coester}}\ and\ \bibinfo {author} {\bibfnamefont {H.}~\bibnamefont {Kümmel}},\ }\href {\doibase 10.1016/0029-5582(60)90140-1} {\bibfield  {journal} {\bibinfo  {journal} {Nucl. Phys.}\ }\textbf {\bibinfo {volume} {17}},\ \bibinfo {pages} {477} (\bibinfo {year} {1960})}\BibitemShut {NoStop}%
\bibitem [{\citenamefont {Kümmel}\ \emph {et~al.}(1978)\citenamefont {Kümmel}, \citenamefont {Lührmann},\ and\ \citenamefont {Zabolitzky}}]{Ref147}%
  \BibitemOpen
  \bibfield  {author} {\bibinfo {author} {\bibfnamefont {H.}~\bibnamefont {Kümmel}}, \bibinfo {author} {\bibfnamefont {K.}~\bibnamefont {Lührmann}}, \ and\ \bibinfo {author} {\bibfnamefont {J.}~\bibnamefont {Zabolitzky}},\ }\href {\doibase 10.1016/0370-1573(78)90081-9} {\bibfield  {journal} {\bibinfo  {journal} {Phys. Rep.}\ }\textbf {\bibinfo {volume} {36}},\ \bibinfo {pages} {1} (\bibinfo {year} {1978})}\BibitemShut {NoStop}%
\bibitem [{\citenamefont {Bishop}(1991)}]{Ref68}%
  \BibitemOpen
  \bibfield  {author} {\bibinfo {author} {\bibfnamefont {R.}~\bibnamefont {Bishop}},\ }\href {\doibase 10.1007/BF01119617} {\bibfield  {journal} {\bibinfo  {journal} {Theor. Chim. Acta}\ }\textbf {\bibinfo {volume} {80}},\ \bibinfo {pages} {95} (\bibinfo {year} {1991})}\BibitemShut {NoStop}%
\bibitem [{\citenamefont {Hagen}\ \emph {et~al.}(2014)\citenamefont {Hagen}, \citenamefont {Papenbrock}, \citenamefont {Hjorth-Jensen},\ and\ \citenamefont {Dean}}]{Ref16}%
  \BibitemOpen
  \bibfield  {author} {\bibinfo {author} {\bibfnamefont {G.}~\bibnamefont {Hagen}}, \bibinfo {author} {\bibfnamefont {T.}~\bibnamefont {Papenbrock}}, \bibinfo {author} {\bibfnamefont {M.}~\bibnamefont {Hjorth-Jensen}}, \ and\ \bibinfo {author} {\bibfnamefont {D.~J.}\ \bibnamefont {Dean}},\ }\href {\doibase 10.1088/0034-4885/77/9/096302} {\bibfield  {journal} {\bibinfo  {journal} {Rep. Prog. Phys.}\ }\textbf {\bibinfo {volume} {77}},\ \bibinfo {pages} {096302} (\bibinfo {year} {2014})}\BibitemShut {NoStop}%
\bibitem [{\citenamefont {Heisenberg}\ and\ \citenamefont {Mihaila}(1999)}]{Ref154}%
  \BibitemOpen
  \bibfield  {author} {\bibinfo {author} {\bibfnamefont {J.~H.}\ \bibnamefont {Heisenberg}}\ and\ \bibinfo {author} {\bibfnamefont {B.}~\bibnamefont {Mihaila}},\ }\href {\doibase 10.1103/PhysRevC.59.1440} {\bibfield  {journal} {\bibinfo  {journal} {Phys. Rev. C}\ }\textbf {\bibinfo {volume} {59}},\ \bibinfo {pages} {1440} (\bibinfo {year} {1999})}\BibitemShut {NoStop}%
\bibitem [{\citenamefont {Dean}\ and\ \citenamefont {Hjorth-Jensen}(2004)}]{Ref148}%
  \BibitemOpen
  \bibfield  {author} {\bibinfo {author} {\bibfnamefont {D.~J.}\ \bibnamefont {Dean}}\ and\ \bibinfo {author} {\bibfnamefont {M.}~\bibnamefont {Hjorth-Jensen}},\ }\href {\doibase 10.1103/PhysRevC.69.054320} {\bibfield  {journal} {\bibinfo  {journal} {Phys. Rev. C}\ }\textbf {\bibinfo {volume} {69}},\ \bibinfo {pages} {054320} (\bibinfo {year} {2004})}\BibitemShut {NoStop}%
\bibitem [{\citenamefont {Bartlett}\ and\ \citenamefont {Musia\l{}}(2007)}]{Ref149}%
  \BibitemOpen
  \bibfield  {author} {\bibinfo {author} {\bibfnamefont {R.~J.}\ \bibnamefont {Bartlett}}\ and\ \bibinfo {author} {\bibfnamefont {M.}~\bibnamefont {Musia\l{}}},\ }\href {\doibase 10.1103/RevModPhys.79.291} {\bibfield  {journal} {\bibinfo  {journal} {Rev. Mod. Phys.}\ }\textbf {\bibinfo {volume} {79}},\ \bibinfo {pages} {291} (\bibinfo {year} {2007})}\BibitemShut {NoStop}%
\bibitem [{\citenamefont {Taube}\ and\ \citenamefont {Bartlett}(2008)}]{Ref140}%
  \BibitemOpen
  \bibfield  {author} {\bibinfo {author} {\bibfnamefont {A.~G.}\ \bibnamefont {Taube}}\ and\ \bibinfo {author} {\bibfnamefont {R.~J.}\ \bibnamefont {Bartlett}},\ }\href {\doibase 10.1063/1.2830236} {\bibfield  {journal} {\bibinfo  {journal} {J. Chem. Phys.}\ }\textbf {\bibinfo {volume} {128}} (\bibinfo {year} {2008}),\ 10.1063/1.2830236},\ \bibinfo {note} {044110}\BibitemShut {NoStop}%
\bibitem [{\citenamefont {Kucharski}\ and\ \citenamefont {Bartlett}(1998)}]{Ref141}%
  \BibitemOpen
  \bibfield  {author} {\bibinfo {author} {\bibfnamefont {S.~A.}\ \bibnamefont {Kucharski}}\ and\ \bibinfo {author} {\bibfnamefont {R.~J.}\ \bibnamefont {Bartlett}},\ }\href {\doibase 10.1063/1.475961} {\bibfield  {journal} {\bibinfo  {journal} {J. Chem. Phys.}\ }\textbf {\bibinfo {volume} {108}},\ \bibinfo {pages} {5243} (\bibinfo {year} {1998})}\BibitemShut {NoStop}%
\bibitem [{\citenamefont {Bartlett}\ \emph {et~al.}(1990)\citenamefont {Bartlett}, \citenamefont {Watts}, \citenamefont {Kucharski},\ and\ \citenamefont {Noga}}]{Ref142}%
  \BibitemOpen
  \bibfield  {author} {\bibinfo {author} {\bibfnamefont {R.~J.}\ \bibnamefont {Bartlett}}, \bibinfo {author} {\bibfnamefont {J.}~\bibnamefont {Watts}}, \bibinfo {author} {\bibfnamefont {S.}~\bibnamefont {Kucharski}}, \ and\ \bibinfo {author} {\bibfnamefont {J.}~\bibnamefont {Noga}},\ }\href {\doibase 10.1016/0009-2614(90)87031-L} {\bibfield  {journal} {\bibinfo  {journal} {Chem. Phys. Lett.}\ }\textbf {\bibinfo {volume} {165}},\ \bibinfo {pages} {513} (\bibinfo {year} {1990})}\BibitemShut {NoStop}%
\bibitem [{\citenamefont {Raghavachari}\ \emph {et~al.}(1989)\citenamefont {Raghavachari}, \citenamefont {Trucks}, \citenamefont {Pople},\ and\ \citenamefont {Head-Gordon}}]{Ref143}%
  \BibitemOpen
  \bibfield  {author} {\bibinfo {author} {\bibfnamefont {K.}~\bibnamefont {Raghavachari}}, \bibinfo {author} {\bibfnamefont {G.~W.}\ \bibnamefont {Trucks}}, \bibinfo {author} {\bibfnamefont {J.~A.}\ \bibnamefont {Pople}}, \ and\ \bibinfo {author} {\bibfnamefont {M.}~\bibnamefont {Head-Gordon}},\ }\href {\doibase 10.1016/S0009-2614(89)87395-6} {\bibfield  {journal} {\bibinfo  {journal} {Chem. Phys. Lett.}\ }\textbf {\bibinfo {volume} {157}},\ \bibinfo {pages} {479} (\bibinfo {year} {1989})}\BibitemShut {NoStop}%
\bibitem [{\citenamefont {Kutzelnigg}(1991)}]{Ref145}%
  \BibitemOpen
  \bibfield  {author} {\bibinfo {author} {\bibfnamefont {W.}~\bibnamefont {Kutzelnigg}},\ }\href {\doibase 10.1007/BF01117418} {\bibfield  {journal} {\bibinfo  {journal} {Theor. Chim. Acta}\ }\textbf {\bibinfo {volume} {80}},\ \bibinfo {pages} {349–386} (\bibinfo {year} {1991})}\BibitemShut {NoStop}%
\bibitem [{\citenamefont {Čižek}\ and\ \citenamefont {Paldus}(1971)}]{Ref150}%
  \BibitemOpen
  \bibfield  {author} {\bibinfo {author} {\bibfnamefont {J.}~\bibnamefont {Čižek}}\ and\ \bibinfo {author} {\bibfnamefont {J.}~\bibnamefont {Paldus}},\ }\href {\doibase 10.1002/qua.560050402} {\bibfield  {journal} {\bibinfo  {journal} {Int. J. Quant. Chem.}\ }\textbf {\bibinfo {volume} {5}},\ \bibinfo {pages} {359} (\bibinfo {year} {1971})}\BibitemShut {NoStop}%
\bibitem [{\citenamefont {He}\ \emph {et~al.}(2001)\citenamefont {He}, \citenamefont {He},\ and\ \citenamefont {Cremer}}]{Ref155}%
  \BibitemOpen
  \bibfield  {author} {\bibinfo {author} {\bibfnamefont {Y.}~\bibnamefont {He}}, \bibinfo {author} {\bibfnamefont {Z.}~\bibnamefont {He}}, \ and\ \bibinfo {author} {\bibfnamefont {D.}~\bibnamefont {Cremer}},\ }\href {\doibase 10.1007/s002140000196} {\bibfield  {journal} {\bibinfo  {journal} {Theoretical Chemistry Accounts: Theory, Computation, and Modeling (Theoretica Chimica Acta)}\ }\textbf {\bibinfo {volume} {105}},\ \bibinfo {pages} {182} (\bibinfo {year} {2001})}\BibitemShut {NoStop}%
\bibitem [{\citenamefont {Margraf}\ and\ \citenamefont {Reuter}(2018)}]{Ref7}%
  \BibitemOpen
  \bibfield  {author} {\bibinfo {author} {\bibfnamefont {J.~T.}\ \bibnamefont {Margraf}}\ and\ \bibinfo {author} {\bibfnamefont {K.}~\bibnamefont {Reuter}},\ }\href {\doibase 10.1021/acs.jpca.8b04455} {\bibfield  {journal} {\bibinfo  {journal} {J. Phys. Chem. A}\ }\textbf {\bibinfo {volume} {122}},\ \bibinfo {pages} {6343} (\bibinfo {year} {2018})},\ \bibinfo {note} {pMID: 29985611}\BibitemShut {NoStop}%
\bibitem [{\citenamefont {Weiler}\ \emph {et~al.}(2022)\citenamefont {Weiler}, \citenamefont {Mihm},\ and\ \citenamefont {Shepherd}}]{Ref67}%
  \BibitemOpen
  \bibfield  {author} {\bibinfo {author} {\bibfnamefont {L.}~\bibnamefont {Weiler}}, \bibinfo {author} {\bibfnamefont {T.~N.}\ \bibnamefont {Mihm}}, \ and\ \bibinfo {author} {\bibfnamefont {J.~J.}\ \bibnamefont {Shepherd}},\ }\href {\doibase 10.1063/5.0086580} {\bibfield  {journal} {\bibinfo  {journal} {J. Chem. Phys.}\ }\textbf {\bibinfo {volume} {156}} (\bibinfo {year} {2022}),\ 10.1063/5.0086580},\ \bibinfo {note} {204109}\BibitemShut {NoStop}%
\bibitem [{\citenamefont {Bishop}\ and\ \citenamefont {Lührmann}(1981)}]{Ref72}%
  \BibitemOpen
  \bibfield  {author} {\bibinfo {author} {\bibfnamefont {R.}~\bibnamefont {Bishop}}\ and\ \bibinfo {author} {\bibfnamefont {K.}~\bibnamefont {Lührmann}},\ }\href {\doibase 10.1016/0378-4363(81)90741-5} {\bibfield  {journal} {\bibinfo  {journal} {Physica B+C}\ }\textbf {\bibinfo {volume} {108}},\ \bibinfo {pages} {873} (\bibinfo {year} {1981})}\BibitemShut {NoStop}%
\bibitem [{\citenamefont {Mihm}\ \emph {et~al.}(2021)\citenamefont {Mihm}, \citenamefont {Yang},\ and\ \citenamefont {Shepherd}}]{Ref74}%
  \BibitemOpen
  \bibfield  {author} {\bibinfo {author} {\bibfnamefont {T.~N.}\ \bibnamefont {Mihm}}, \bibinfo {author} {\bibfnamefont {B.}~\bibnamefont {Yang}}, \ and\ \bibinfo {author} {\bibfnamefont {J.~J.}\ \bibnamefont {Shepherd}},\ }\href {\doibase 10.1021/acs.jctc.0c01171} {\bibfield  {journal} {\bibinfo  {journal} {J. Chem. Theor. Comp.}\ }\textbf {\bibinfo {volume} {17}},\ \bibinfo {pages} {2752} (\bibinfo {year} {2021})},\ \bibinfo {note} {pMID: 33830754}\BibitemShut {NoStop}%
\bibitem [{\citenamefont {Drischler}\ \emph {et~al.}(2019)\citenamefont {Drischler}, \citenamefont {Hebeler},\ and\ \citenamefont {Schwenk}}]{Drischler2019}%
  \BibitemOpen
  \bibfield  {author} {\bibinfo {author} {\bibfnamefont {C.}~\bibnamefont {Drischler}}, \bibinfo {author} {\bibfnamefont {K.}~\bibnamefont {Hebeler}}, \ and\ \bibinfo {author} {\bibfnamefont {A.}~\bibnamefont {Schwenk}},\ }\href {\doibase 10.1103/PhysRevLett.122.042501} {\bibfield  {journal} {\bibinfo  {journal} {Phys. Rev. Lett.}\ }\textbf {\bibinfo {volume} {122}},\ \bibinfo {pages} {042501} (\bibinfo {year} {2019})}\BibitemShut {NoStop}%
\bibitem [{\citenamefont {Drischler}\ \emph {et~al.}(2021{\natexlab{a}})\citenamefont {Drischler}, \citenamefont {Holt},\ and\ \citenamefont {Wellenhofer}}]{Drischler2021}%
  \BibitemOpen
  \bibfield  {author} {\bibinfo {author} {\bibfnamefont {C.}~\bibnamefont {Drischler}}, \bibinfo {author} {\bibfnamefont {J.}~\bibnamefont {Holt}}, \ and\ \bibinfo {author} {\bibfnamefont {C.}~\bibnamefont {Wellenhofer}},\ }\href {\doibase 10.1146/annurev-nucl-102419-041903} {\bibfield  {journal} {\bibinfo  {journal} {Annual Review of Nuclear and Particle Science}\ }\textbf {\bibinfo {volume} {71}},\ \bibinfo {pages} {403} (\bibinfo {year} {2021}{\natexlab{a}})}\BibitemShut {NoStop}%
\bibitem [{\citenamefont {Carbone}\ \emph {et~al.}(2013)\citenamefont {Carbone}, \citenamefont {Polls},\ and\ \citenamefont {Rios}}]{Ref50}%
  \BibitemOpen
  \bibfield  {author} {\bibinfo {author} {\bibfnamefont {A.}~\bibnamefont {Carbone}}, \bibinfo {author} {\bibfnamefont {A.}~\bibnamefont {Polls}}, \ and\ \bibinfo {author} {\bibfnamefont {A.}~\bibnamefont {Rios}},\ }\href {\doibase 10.1103/PhysRevC.88.044302} {\bibfield  {journal} {\bibinfo  {journal} {Phys. Rev. C}\ }\textbf {\bibinfo {volume} {88}},\ \bibinfo {pages} {044302} (\bibinfo {year} {2013})}\BibitemShut {NoStop}%
\bibitem [{\citenamefont {Barbieri}\ and\ \citenamefont {Carbone}(2017)}]{Barbieri2017}%
  \BibitemOpen
  \bibfield  {author} {\bibinfo {author} {\bibfnamefont {C.}~\bibnamefont {Barbieri}}\ and\ \bibinfo {author} {\bibfnamefont {A.}~\bibnamefont {Carbone}},\ }\enquote {\bibinfo {title} {Self-consistent green's function approaches},}\ in\ \href {\doibase 10.1007/978-3-319-53336-0_11} {\emph {\bibinfo {booktitle} {An Advanced Course in Computational Nuclear Physics: Bridging the Scales from Quarks to Neutron Stars}}},\ \bibinfo {editor} {edited by\ \bibinfo {editor} {\bibfnamefont {M.}~\bibnamefont {Hjorth-Jensen}}, \bibinfo {editor} {\bibfnamefont {M.~P.}\ \bibnamefont {Lombardo}}, \ and\ \bibinfo {editor} {\bibfnamefont {U.}~\bibnamefont {van Kolck}}}\ (\bibinfo  {publisher} {Springer International Publishing},\ \bibinfo {address} {Cham},\ \bibinfo {year} {2017})\ pp.\ \bibinfo {pages} {571--644}\BibitemShut {NoStop}%
\bibitem [{\citenamefont {Hergert}\ \emph {et~al.}(2017)\citenamefont {Hergert}, \citenamefont {Bogner}, \citenamefont {Lietz}, \citenamefont {Morris}, \citenamefont {Novario}, \citenamefont {Parzuchowski},\ and\ \citenamefont {Yuan}}]{Hergert2017}%
  \BibitemOpen
  \bibfield  {author} {\bibinfo {author} {\bibfnamefont {H.}~\bibnamefont {Hergert}}, \bibinfo {author} {\bibfnamefont {S.~K.}\ \bibnamefont {Bogner}}, \bibinfo {author} {\bibfnamefont {J.~G.}\ \bibnamefont {Lietz}}, \bibinfo {author} {\bibfnamefont {T.~D.}\ \bibnamefont {Morris}}, \bibinfo {author} {\bibfnamefont {S.~J.}\ \bibnamefont {Novario}}, \bibinfo {author} {\bibfnamefont {N.~M.}\ \bibnamefont {Parzuchowski}}, \ and\ \bibinfo {author} {\bibfnamefont {F.}~\bibnamefont {Yuan}},\ }\enquote {\bibinfo {title} {In-medium similarity renormalization group approach to the nuclear many-body problem},}\ in\ \href {\doibase 10.1007/978-3-319-53336-0_10} {\emph {\bibinfo {booktitle} {An Advanced Course in Computational Nuclear Physics: Bridging the Scales from Quarks to Neutron Stars}}},\ \bibinfo {editor} {edited by\ \bibinfo {editor} {\bibfnamefont {M.}~\bibnamefont {Hjorth-Jensen}}, \bibinfo {editor} {\bibfnamefont {M.~P.}\ \bibnamefont {Lombardo}}, \ and\ \bibinfo {editor} {\bibfnamefont {U.}~\bibnamefont {van
  Kolck}}}\ (\bibinfo  {publisher} {Springer International Publishing},\ \bibinfo {address} {Cham},\ \bibinfo {year} {2017})\ pp.\ \bibinfo {pages} {477--570}\BibitemShut {NoStop}%
\bibitem [{\citenamefont {Marino}\ \emph {et~al.}(2024)\citenamefont {Marino}, \citenamefont {Jiang},\ and\ \citenamefont {Novario}}]{Marino2024}%
  \BibitemOpen
  \bibfield  {author} {\bibinfo {author} {\bibfnamefont {F.}~\bibnamefont {Marino}}, \bibinfo {author} {\bibfnamefont {W.}~\bibnamefont {Jiang}}, \ and\ \bibinfo {author} {\bibfnamefont {S.~J.}\ \bibnamefont {Novario}},\ }\href {https://arxiv.org/abs/2407.17098} {\enquote {\bibinfo {title} {Diagrammatic ab initio methods for infinite nuclear matter with modern chiral interactions},}\ } (\bibinfo {year} {2024}),\ \Eprint {http://arxiv.org/abs/2407.17098} {arXiv:2407.17098 [nucl-th]} \BibitemShut {NoStop}%
\bibitem [{\citenamefont {Raghavachari}\ \emph {et~al.}(2013)\citenamefont {Raghavachari}, \citenamefont {Trucks}, \citenamefont {Pople},\ and\ \citenamefont {Head-Gordon}}]{Ref158}%
  \BibitemOpen
  \bibfield  {author} {\bibinfo {author} {\bibfnamefont {K.}~\bibnamefont {Raghavachari}}, \bibinfo {author} {\bibfnamefont {G.}~\bibnamefont {Trucks}}, \bibinfo {author} {\bibfnamefont {J.}~\bibnamefont {Pople}}, \ and\ \bibinfo {author} {\bibfnamefont {M.}~\bibnamefont {Head-Gordon}},\ }\href {\doibase 10.1016/j.cplett.2013.08.064} {\bibfield  {journal} {\bibinfo  {journal} {Chem. Phys. Lett.}\ }\textbf {\bibinfo {volume} {589}},\ \bibinfo {pages} {37} (\bibinfo {year} {2013})}\BibitemShut {NoStop}%
\bibitem [{\citenamefont {Shavitt}\ and\ \citenamefont {Bartlett}(2009)}]{Ref21}%
  \BibitemOpen
  \bibfield  {author} {\bibinfo {author} {\bibfnamefont {I.}~\bibnamefont {Shavitt}}\ and\ \bibinfo {author} {\bibfnamefont {R.~J.}\ \bibnamefont {Bartlett}},\ }\href {\doibase 10.1017/CBO9780511596834} {\emph {\bibinfo {title} {Many-Body Methods in Chemistry and Physics: MBPT and Coupled-Cluster Theory}}},\ Cambridge Molecular Science\ (\bibinfo  {publisher} {Cambridge University Press},\ \bibinfo {year} {2009})\BibitemShut {NoStop}%
\bibitem [{\citenamefont {Urban}\ \emph {et~al.}(1985)\citenamefont {Urban}, \citenamefont {Noga}, \citenamefont {Cole},\ and\ \citenamefont {Bartlett}}]{Ref157}%
  \BibitemOpen
  \bibfield  {author} {\bibinfo {author} {\bibfnamefont {M.}~\bibnamefont {Urban}}, \bibinfo {author} {\bibfnamefont {J.}~\bibnamefont {Noga}}, \bibinfo {author} {\bibfnamefont {S.~J.}\ \bibnamefont {Cole}}, \ and\ \bibinfo {author} {\bibfnamefont {R.~J.}\ \bibnamefont {Bartlett}},\ }\href {\doibase 10.1063/1.449067} {\bibfield  {journal} {\bibinfo  {journal} {J. Chem. Phys.}\ }\textbf {\bibinfo {volume} {83}},\ \bibinfo {pages} {4041–4046} (\bibinfo {year} {1985})}\BibitemShut {NoStop}%
\bibitem [{\citenamefont {Boehnlein}\ \emph {et~al.}(2022)\citenamefont {Boehnlein}, \citenamefont {Diefenthaler}, \citenamefont {Sato}, \citenamefont {Schram}, \citenamefont {Ziegler}, \citenamefont {Fanelli}, \citenamefont {Hjorth-Jensen}, \citenamefont {Horn}, \citenamefont {Kuchera}, \citenamefont {Lee}, \citenamefont {Nazarewicz}, \citenamefont {Ostroumov}, \citenamefont {Orginos}, \citenamefont {Poon}, \citenamefont {Wang}, \citenamefont {Scheinker}, \citenamefont {Smith},\ and\ \citenamefont {Pang}}]{Ref210}%
  \BibitemOpen
  \bibfield  {author} {\bibinfo {author} {\bibfnamefont {A.}~\bibnamefont {Boehnlein}}, \bibinfo {author} {\bibfnamefont {M.}~\bibnamefont {Diefenthaler}}, \bibinfo {author} {\bibfnamefont {N.}~\bibnamefont {Sato}}, \bibinfo {author} {\bibfnamefont {M.}~\bibnamefont {Schram}}, \bibinfo {author} {\bibfnamefont {V.}~\bibnamefont {Ziegler}}, \bibinfo {author} {\bibfnamefont {C.}~\bibnamefont {Fanelli}}, \bibinfo {author} {\bibfnamefont {M.}~\bibnamefont {Hjorth-Jensen}}, \bibinfo {author} {\bibfnamefont {T.}~\bibnamefont {Horn}}, \bibinfo {author} {\bibfnamefont {M.~P.}\ \bibnamefont {Kuchera}}, \bibinfo {author} {\bibfnamefont {D.}~\bibnamefont {Lee}}, \bibinfo {author} {\bibfnamefont {W.}~\bibnamefont {Nazarewicz}}, \bibinfo {author} {\bibfnamefont {P.}~\bibnamefont {Ostroumov}}, \bibinfo {author} {\bibfnamefont {K.}~\bibnamefont {Orginos}}, \bibinfo {author} {\bibfnamefont {A.}~\bibnamefont {Poon}}, \bibinfo {author} {\bibfnamefont {X.-N.}\ \bibnamefont {Wang}}, \bibinfo {author} {\bibfnamefont
  {A.}~\bibnamefont {Scheinker}}, \bibinfo {author} {\bibfnamefont {M.~S.}\ \bibnamefont {Smith}}, \ and\ \bibinfo {author} {\bibfnamefont {L.-G.}\ \bibnamefont {Pang}},\ }\href {\doibase 10.1103/RevModPhys.94.031003} {\bibfield  {journal} {\bibinfo  {journal} {Rev. Mod. Phys.}\ }\textbf {\bibinfo {volume} {94}},\ \bibinfo {pages} {031003} (\bibinfo {year} {2022})}\BibitemShut {NoStop}%
\bibitem [{\citenamefont {Carleo}\ \emph {et~al.}(2019)\citenamefont {Carleo}, \citenamefont {Cirac}, \citenamefont {Cranmer}, \citenamefont {Daudet}, \citenamefont {Schuld}, \citenamefont {Tishby}, \citenamefont {Vogt-Maranto},\ and\ \citenamefont {Zdeborov\'a}}]{Ref208}%
  \BibitemOpen
  \bibfield  {author} {\bibinfo {author} {\bibfnamefont {G.}~\bibnamefont {Carleo}}, \bibinfo {author} {\bibfnamefont {I.}~\bibnamefont {Cirac}}, \bibinfo {author} {\bibfnamefont {K.}~\bibnamefont {Cranmer}}, \bibinfo {author} {\bibfnamefont {L.}~\bibnamefont {Daudet}}, \bibinfo {author} {\bibfnamefont {M.}~\bibnamefont {Schuld}}, \bibinfo {author} {\bibfnamefont {N.}~\bibnamefont {Tishby}}, \bibinfo {author} {\bibfnamefont {L.}~\bibnamefont {Vogt-Maranto}}, \ and\ \bibinfo {author} {\bibfnamefont {L.}~\bibnamefont {Zdeborov\'a}},\ }\href {\doibase 10.1103/RevModPhys.91.045002} {\bibfield  {journal} {\bibinfo  {journal} {Rev. Mod. Phys.}\ }\textbf {\bibinfo {volume} {91}},\ \bibinfo {pages} {045002} (\bibinfo {year} {2019})}\BibitemShut {NoStop}%
\bibitem [{\citenamefont {Mehta}\ \emph {et~al.}(2019)\citenamefont {Mehta}, \citenamefont {Bukov}, \citenamefont {Wang}, \citenamefont {Day}, \citenamefont {Richardson}, \citenamefont {Fisher},\ and\ \citenamefont {Schwab}}]{Ref209}%
  \BibitemOpen
  \bibfield  {author} {\bibinfo {author} {\bibfnamefont {P.}~\bibnamefont {Mehta}}, \bibinfo {author} {\bibfnamefont {M.}~\bibnamefont {Bukov}}, \bibinfo {author} {\bibfnamefont {C.}~\bibnamefont {Wang}}, \bibinfo {author} {\bibfnamefont {A.~G.}\ \bibnamefont {Day}}, \bibinfo {author} {\bibfnamefont {C.}~\bibnamefont {Richardson}}, \bibinfo {author} {\bibfnamefont {C.~K.}\ \bibnamefont {Fisher}}, \ and\ \bibinfo {author} {\bibfnamefont {D.~J.}\ \bibnamefont {Schwab}},\ }\href {\doibase 10.1016/j.physrep.2019.03.001} {\bibfield  {journal} {\bibinfo  {journal} {Phys. Rep.}\ }\textbf {\bibinfo {volume} {810}},\ \bibinfo {pages} {1 } (\bibinfo {year} {2019})}\BibitemShut {NoStop}%
\bibitem [{\citenamefont {Clark}\ \emph {et~al.}()\citenamefont {Clark}, \citenamefont {Mavrommatis}, \citenamefont {Athanassopoulos}, \citenamefont {Dakos},\ and\ \citenamefont {Gernoth}}]{Ref24}%
  \BibitemOpen
  \bibfield  {author} {\bibinfo {author} {\bibfnamefont {J.~W.}\ \bibnamefont {Clark}}, \bibinfo {author} {\bibfnamefont {E.}~\bibnamefont {Mavrommatis}}, \bibinfo {author} {\bibfnamefont {S.}~\bibnamefont {Athanassopoulos}}, \bibinfo {author} {\bibfnamefont {A.}~\bibnamefont {Dakos}}, \ and\ \bibinfo {author} {\bibfnamefont {K.}~\bibnamefont {Gernoth}},\ }\enquote {\bibinfo {title} {Statistical modeling of nuclear systematics},}\ in\ \href {\doibase 10.1142/9789812811127_0008} {\emph {\bibinfo {booktitle} {Fission Dynamics of Atomic Clusters and Nuclei}}},\ pp.\ \bibinfo {pages} {76--85}\BibitemShut {NoStop}%
\bibitem [{\citenamefont {Athanassopoulos}\ \emph {et~al.}(2004)\citenamefont {Athanassopoulos}, \citenamefont {Mavrommatis}, \citenamefont {Gernoth},\ and\ \citenamefont {Clark}}]{Ref25}%
  \BibitemOpen
  \bibfield  {author} {\bibinfo {author} {\bibfnamefont {S.}~\bibnamefont {Athanassopoulos}}, \bibinfo {author} {\bibfnamefont {E.}~\bibnamefont {Mavrommatis}}, \bibinfo {author} {\bibfnamefont {K.}~\bibnamefont {Gernoth}}, \ and\ \bibinfo {author} {\bibfnamefont {J.}~\bibnamefont {Clark}},\ }\href {\doibase https://doi.org/10.1016/j.nuclphysa.2004.08.006} {\bibfield  {journal} {\bibinfo  {journal} {Nucl. Phys. A}\ }\textbf {\bibinfo {volume} {743}},\ \bibinfo {pages} {222} (\bibinfo {year} {2004})}\BibitemShut {NoStop}%
\bibitem [{\citenamefont {Costiris}\ \emph {et~al.}(2009)\citenamefont {Costiris}, \citenamefont {Mavrommatis}, \citenamefont {Gernoth},\ and\ \citenamefont {Clark}}]{Ref26}%
  \BibitemOpen
  \bibfield  {author} {\bibinfo {author} {\bibfnamefont {N.~J.}\ \bibnamefont {Costiris}}, \bibinfo {author} {\bibfnamefont {E.}~\bibnamefont {Mavrommatis}}, \bibinfo {author} {\bibfnamefont {K.~A.}\ \bibnamefont {Gernoth}}, \ and\ \bibinfo {author} {\bibfnamefont {J.~W.}\ \bibnamefont {Clark}},\ }\href {\doibase 10.1103/PhysRevC.80.044332} {\bibfield  {journal} {\bibinfo  {journal} {Phys. Rev. C}\ }\textbf {\bibinfo {volume} {80}},\ \bibinfo {pages} {044332} (\bibinfo {year} {2009})}\BibitemShut {NoStop}%
\bibitem [{\citenamefont {Akkoyun}\ \emph {et~al.}(2013)\citenamefont {Akkoyun}, \citenamefont {Bayram}, \citenamefont {Kara},\ and\ \citenamefont {Sinan}}]{Ref27}%
  \BibitemOpen
  \bibfield  {author} {\bibinfo {author} {\bibfnamefont {S.}~\bibnamefont {Akkoyun}}, \bibinfo {author} {\bibfnamefont {T.}~\bibnamefont {Bayram}}, \bibinfo {author} {\bibfnamefont {S.~O.}\ \bibnamefont {Kara}}, \ and\ \bibinfo {author} {\bibfnamefont {A.}~\bibnamefont {Sinan}},\ }\href {\doibase 10.1088/0954-3899/40/5/055106} {\bibfield  {journal} {\bibinfo  {journal} {J. Phys. G}\ }\textbf {\bibinfo {volume} {40}},\ \bibinfo {pages} {055106} (\bibinfo {year} {2013})}\BibitemShut {NoStop}%
\bibitem [{\citenamefont {Utama}\ \emph {et~al.}(2016{\natexlab{a}})\citenamefont {Utama}, \citenamefont {Piekarewicz},\ and\ \citenamefont {Prosper}}]{Ref28}%
  \BibitemOpen
  \bibfield  {author} {\bibinfo {author} {\bibfnamefont {R.}~\bibnamefont {Utama}}, \bibinfo {author} {\bibfnamefont {J.}~\bibnamefont {Piekarewicz}}, \ and\ \bibinfo {author} {\bibfnamefont {H.~B.}\ \bibnamefont {Prosper}},\ }\href {\doibase 10.1103/PhysRevC.93.014311} {\bibfield  {journal} {\bibinfo  {journal} {Phys. Rev. C}\ }\textbf {\bibinfo {volume} {93}},\ \bibinfo {pages} {014311} (\bibinfo {year} {2016}{\natexlab{a}})}\BibitemShut {NoStop}%
\bibitem [{\citenamefont {Utama}\ \emph {et~al.}(2016{\natexlab{b}})\citenamefont {Utama}, \citenamefont {Chen},\ and\ \citenamefont {Piekarewicz}}]{Ref29}%
  \BibitemOpen
  \bibfield  {author} {\bibinfo {author} {\bibfnamefont {R.}~\bibnamefont {Utama}}, \bibinfo {author} {\bibfnamefont {W.-C.}\ \bibnamefont {Chen}}, \ and\ \bibinfo {author} {\bibfnamefont {J.}~\bibnamefont {Piekarewicz}},\ }\href {\doibase 10.1088/0954-3899/43/11/114002} {\bibfield  {journal} {\bibinfo  {journal} {J. Phys. G:}\ }\textbf {\bibinfo {volume} {43}},\ \bibinfo {pages} {114002} (\bibinfo {year} {2016}{\natexlab{b}})}\BibitemShut {NoStop}%
\bibitem [{\citenamefont {Utama}\ and\ \citenamefont {Piekarewicz}(2017)}]{Ref30}%
  \BibitemOpen
  \bibfield  {author} {\bibinfo {author} {\bibfnamefont {R.}~\bibnamefont {Utama}}\ and\ \bibinfo {author} {\bibfnamefont {J.}~\bibnamefont {Piekarewicz}},\ }\href {\doibase 10.1103/PhysRevC.96.044308} {\bibfield  {journal} {\bibinfo  {journal} {Phys. Rev. C}\ }\textbf {\bibinfo {volume} {96}},\ \bibinfo {pages} {044308} (\bibinfo {year} {2017})}\BibitemShut {NoStop}%
\bibitem [{\citenamefont {Utama}\ and\ \citenamefont {Piekarewicz}(2018)}]{Ref31}%
  \BibitemOpen
  \bibfield  {author} {\bibinfo {author} {\bibfnamefont {R.}~\bibnamefont {Utama}}\ and\ \bibinfo {author} {\bibfnamefont {J.}~\bibnamefont {Piekarewicz}},\ }\href {\doibase 10.1103/PhysRevC.97.014306} {\bibfield  {journal} {\bibinfo  {journal} {Phys. Rev. C}\ }\textbf {\bibinfo {volume} {97}},\ \bibinfo {pages} {014306} (\bibinfo {year} {2018})}\BibitemShut {NoStop}%
\bibitem [{\citenamefont {Carleo}\ and\ \citenamefont {Troyer}(2017)}]{Ref32}%
  \BibitemOpen
  \bibfield  {author} {\bibinfo {author} {\bibfnamefont {G.}~\bibnamefont {Carleo}}\ and\ \bibinfo {author} {\bibfnamefont {M.}~\bibnamefont {Troyer}},\ }\href {\doibase 10.1126/science.aag2302} {\bibfield  {journal} {\bibinfo  {journal} {Science}\ }\textbf {\bibinfo {volume} {355}},\ \bibinfo {pages} {602} (\bibinfo {year} {2017})}\BibitemShut {NoStop}%
\bibitem [{\citenamefont {Adams}\ \emph {et~al.}(2021)\citenamefont {Adams}, \citenamefont {Carleo}, \citenamefont {Lovato},\ and\ \citenamefont {Rocco}}]{Ref211}%
  \BibitemOpen
  \bibfield  {author} {\bibinfo {author} {\bibfnamefont {C.}~\bibnamefont {Adams}}, \bibinfo {author} {\bibfnamefont {G.}~\bibnamefont {Carleo}}, \bibinfo {author} {\bibfnamefont {A.}~\bibnamefont {Lovato}}, \ and\ \bibinfo {author} {\bibfnamefont {N.}~\bibnamefont {Rocco}},\ }\href {\doibase 10.1103/PhysRevLett.127.022502} {\bibfield  {journal} {\bibinfo  {journal} {Phys. Rev. Lett.}\ }\textbf {\bibinfo {volume} {127}},\ \bibinfo {pages} {022502} (\bibinfo {year} {2021})}\BibitemShut {NoStop}%
\bibitem [{\citenamefont {Lovato}\ \emph {et~al.}(2022)\citenamefont {Lovato}, \citenamefont {Adams}, \citenamefont {Carleo},\ and\ \citenamefont {Rocco}}]{Ref212}%
  \BibitemOpen
  \bibfield  {author} {\bibinfo {author} {\bibfnamefont {A.}~\bibnamefont {Lovato}}, \bibinfo {author} {\bibfnamefont {C.}~\bibnamefont {Adams}}, \bibinfo {author} {\bibfnamefont {G.}~\bibnamefont {Carleo}}, \ and\ \bibinfo {author} {\bibfnamefont {N.}~\bibnamefont {Rocco}},\ }\href {\doibase 10.1103/PhysRevResearch.4.043178} {\bibfield  {journal} {\bibinfo  {journal} {Phys. Rev. Res.}\ }\textbf {\bibinfo {volume} {4}},\ \bibinfo {pages} {043178} (\bibinfo {year} {2022})}\BibitemShut {NoStop}%
\bibitem [{\citenamefont {Pescia}\ \emph {et~al.}(2022)\citenamefont {Pescia}, \citenamefont {Han}, \citenamefont {Lovato}, \citenamefont {Lu},\ and\ \citenamefont {Carleo}}]{Ref213}%
  \BibitemOpen
  \bibfield  {author} {\bibinfo {author} {\bibfnamefont {G.}~\bibnamefont {Pescia}}, \bibinfo {author} {\bibfnamefont {J.}~\bibnamefont {Han}}, \bibinfo {author} {\bibfnamefont {A.}~\bibnamefont {Lovato}}, \bibinfo {author} {\bibfnamefont {J.}~\bibnamefont {Lu}}, \ and\ \bibinfo {author} {\bibfnamefont {G.}~\bibnamefont {Carleo}},\ }\href {\doibase 10.1103/PhysRevResearch.4.023138} {\bibfield  {journal} {\bibinfo  {journal} {Phys. Rev. Res.}\ }\textbf {\bibinfo {volume} {4}},\ \bibinfo {pages} {023138} (\bibinfo {year} {2022})}\BibitemShut {NoStop}%
\bibitem [{\citenamefont {Kessler}\ \emph {et~al.}(2021)\citenamefont {Kessler}, \citenamefont {Calcavecchia},\ and\ \citenamefont {Kühne}}]{Ref214}%
  \BibitemOpen
  \bibfield  {author} {\bibinfo {author} {\bibfnamefont {J.}~\bibnamefont {Kessler}}, \bibinfo {author} {\bibfnamefont {F.}~\bibnamefont {Calcavecchia}}, \ and\ \bibinfo {author} {\bibfnamefont {T.~D.}\ \bibnamefont {Kühne}},\ }\href {\doibase 10.1002/adts.202000269} {\bibfield  {journal} {\bibinfo  {journal} {Adv. Theor. Sim.}\ }\textbf {\bibinfo {volume} {4}},\ \bibinfo {pages} {2000269} (\bibinfo {year} {2021})}\BibitemShut {NoStop}%
\bibitem [{\citenamefont {Butler}\ \emph {et~al.}(2024)\citenamefont {Butler}, \citenamefont {Hjorth-Jensen},\ and\ \citenamefont {Leitz}}]{electron_gas_paper}%
  \BibitemOpen
  \bibfield  {author} {\bibinfo {author} {\bibfnamefont {J.}~\bibnamefont {Butler}}, \bibinfo {author} {\bibfnamefont {M.}~\bibnamefont {Hjorth-Jensen}}, \ and\ \bibinfo {author} {\bibfnamefont {J.}~\bibnamefont {Leitz}},\ }\href@noop {} {\bibfield  {journal} {\bibinfo  {journal} {J. Chem. Phys.}\ }\textbf {\bibinfo {volume} {161}},\ \bibinfo {pages} {in press} (\bibinfo {year} {2024})}\BibitemShut {NoStop}%
\bibitem [{\citenamefont {Drischler}\ \emph {et~al.}(2021{\natexlab{b}})\citenamefont {Drischler}, \citenamefont {Holt},\ and\ \citenamefont {Wellenhofer}}]{Ref215}%
  \BibitemOpen
  \bibfield  {author} {\bibinfo {author} {\bibfnamefont {C.}~\bibnamefont {Drischler}}, \bibinfo {author} {\bibfnamefont {J.}~\bibnamefont {Holt}}, \ and\ \bibinfo {author} {\bibfnamefont {C.}~\bibnamefont {Wellenhofer}},\ }\href {\doibase 10.1146/annurev-nucl-102419-041903} {\bibfield  {journal} {\bibinfo  {journal} {Ann. Rev. Nucl. Part. Sci.}\ }\textbf {\bibinfo {volume} {71}},\ \bibinfo {pages} {403} (\bibinfo {year} {2021}{\natexlab{b}})}\BibitemShut {NoStop}%
\bibitem [{\citenamefont {Drischler}\ \emph {et~al.}(2021{\natexlab{c}})\citenamefont {Drischler}, \citenamefont {Han}, \citenamefont {Lattimer}, \citenamefont {Prakash}, \citenamefont {Reddy},\ and\ \citenamefont {Zhao}}]{Ref216}%
  \BibitemOpen
  \bibfield  {author} {\bibinfo {author} {\bibfnamefont {C.}~\bibnamefont {Drischler}}, \bibinfo {author} {\bibfnamefont {S.}~\bibnamefont {Han}}, \bibinfo {author} {\bibfnamefont {J.~M.}\ \bibnamefont {Lattimer}}, \bibinfo {author} {\bibfnamefont {M.}~\bibnamefont {Prakash}}, \bibinfo {author} {\bibfnamefont {S.}~\bibnamefont {Reddy}}, \ and\ \bibinfo {author} {\bibfnamefont {T.}~\bibnamefont {Zhao}},\ }\href {\doibase 10.1103/PhysRevC.103.045808} {\bibfield  {journal} {\bibinfo  {journal} {Phys. Rev. C}\ }\textbf {\bibinfo {volume} {103}},\ \bibinfo {pages} {045808} (\bibinfo {year} {2021}{\natexlab{c}})}\BibitemShut {NoStop}%
\bibitem [{\citenamefont {Burgio}\ \emph {et~al.}(2021)\citenamefont {Burgio}, \citenamefont {Schulze}, \citenamefont {Vidaña},\ and\ \citenamefont {Wei}}]{Ref217}%
  \BibitemOpen
  \bibfield  {author} {\bibinfo {author} {\bibfnamefont {G.}~\bibnamefont {Burgio}}, \bibinfo {author} {\bibfnamefont {H.-J.}\ \bibnamefont {Schulze}}, \bibinfo {author} {\bibfnamefont {I.}~\bibnamefont {Vidaña}}, \ and\ \bibinfo {author} {\bibfnamefont {J.-B.}\ \bibnamefont {Wei}},\ }\href {\doibase 10.1016/j.ppnp.2021.103879} {\bibfield  {journal} {\bibinfo  {journal} {Prog. Part. Nucl. Phys.}\ }\textbf {\bibinfo {volume} {120}},\ \bibinfo {pages} {103879} (\bibinfo {year} {2021})}\BibitemShut {NoStop}%
\bibitem [{\citenamefont {Baldo}\ and\ \citenamefont {Burgio}(2016)}]{symE1}%
  \BibitemOpen
  \bibfield  {author} {\bibinfo {author} {\bibfnamefont {M.}~\bibnamefont {Baldo}}\ and\ \bibinfo {author} {\bibfnamefont {G.}~\bibnamefont {Burgio}},\ }\href {\doibase https://doi.org/10.1016/j.ppnp.2016.06.006} {\bibfield  {journal} {\bibinfo  {journal} {Progress in Particle and Nuclear Physics}\ }\textbf {\bibinfo {volume} {91}},\ \bibinfo {pages} {203} (\bibinfo {year} {2016})}\BibitemShut {NoStop}%
\bibitem [{\citenamefont {Lattimer}(2014)}]{symE2}%
  \BibitemOpen
  \bibfield  {author} {\bibinfo {author} {\bibfnamefont {J.~M.}\ \bibnamefont {Lattimer}},\ }\href {\doibase https://doi.org/10.1016/j.nuclphysa.2014.04.008} {\bibfield  {journal} {\bibinfo  {journal} {Nuclear Physics A}\ }\textbf {\bibinfo {volume} {928}},\ \bibinfo {pages} {276} (\bibinfo {year} {2014})},\ \bibinfo {note} {special Issue Dedicated to the Memory of Gerald E Brown (1926-2013)}\BibitemShut {NoStop}%
\bibitem [{\citenamefont {Ekstr\"om}\ \emph {et~al.}(2018)\citenamefont {Ekstr\"om}, \citenamefont {Hagen}, \citenamefont {Morris}, \citenamefont {Papenbrock},\ and\ \citenamefont {Schwartz}}]{Ref200}%
  \BibitemOpen
  \bibfield  {author} {\bibinfo {author} {\bibfnamefont {A.}~\bibnamefont {Ekstr\"om}}, \bibinfo {author} {\bibfnamefont {G.}~\bibnamefont {Hagen}}, \bibinfo {author} {\bibfnamefont {T.~D.}\ \bibnamefont {Morris}}, \bibinfo {author} {\bibfnamefont {T.}~\bibnamefont {Papenbrock}}, \ and\ \bibinfo {author} {\bibfnamefont {P.~D.}\ \bibnamefont {Schwartz}},\ }\href {\doibase 10.1103/PhysRevC.97.024332} {\bibfield  {journal} {\bibinfo  {journal} {Phys. Rev. C}\ }\textbf {\bibinfo {volume} {97}},\ \bibinfo {pages} {024332} (\bibinfo {year} {2018})}\BibitemShut {NoStop}%
\bibitem [{\citenamefont {Jiang}\ \emph {et~al.}(2020)\citenamefont {Jiang}, \citenamefont {Ekstr\"om}, \citenamefont {Forss\'en}, \citenamefont {Hagen}, \citenamefont {Jansen},\ and\ \citenamefont {Papenbrock}}]{Ref201}%
  \BibitemOpen
  \bibfield  {author} {\bibinfo {author} {\bibfnamefont {W.~G.}\ \bibnamefont {Jiang}}, \bibinfo {author} {\bibfnamefont {A.}~\bibnamefont {Ekstr\"om}}, \bibinfo {author} {\bibfnamefont {C.}~\bibnamefont {Forss\'en}}, \bibinfo {author} {\bibfnamefont {G.}~\bibnamefont {Hagen}}, \bibinfo {author} {\bibfnamefont {G.~R.}\ \bibnamefont {Jansen}}, \ and\ \bibinfo {author} {\bibfnamefont {T.}~\bibnamefont {Papenbrock}},\ }\href {\doibase 10.1103/PhysRevC.102.054301} {\bibfield  {journal} {\bibinfo  {journal} {Phys. Rev. C}\ }\textbf {\bibinfo {volume} {102}},\ \bibinfo {pages} {054301} (\bibinfo {year} {2020})}\BibitemShut {NoStop}%
\bibitem [{\citenamefont {Butler}(2023)}]{my_thesis}%
  \BibitemOpen
  \bibfield  {author} {\bibinfo {author} {\bibfnamefont {J.~L.}\ \bibnamefont {Butler}},\ }\href@noop {} {\enquote {\bibinfo {title} {Machine learning and coupled cluster theory applied to infinite matter},}\ } (\bibinfo {year} {2023})\BibitemShut {NoStop}%
\bibitem [{\citenamefont {Murphy}(2012)}]{Ref218}%
  \BibitemOpen
  \bibfield  {author} {\bibinfo {author} {\bibfnamefont {K.~P.}\ \bibnamefont {Murphy}},\ }\href@noop {} {\emph {\bibinfo {title} {Machine Learning: A Probabilistic Perspective}}}\ (\bibinfo  {publisher} {The MIT Press, Cambdridge, Massachusetts},\ \bibinfo {year} {2012})\BibitemShut {NoStop}%
\bibitem [{\citenamefont {Rasmussen}\ and\ \citenamefont {Williams}(2006)}]{Ref219}%
  \BibitemOpen
  \bibfield  {author} {\bibinfo {author} {\bibfnamefont {C.~E.}\ \bibnamefont {Rasmussen}}\ and\ \bibinfo {author} {\bibfnamefont {C.~K.~I.}\ \bibnamefont {Williams}},\ }\href@noop {} {\emph {\bibinfo {title} {Gaussian Processes for Machine Learning}}}\ (\bibinfo  {publisher} {The MIT Press, Cambdridge, Massachusetts},\ \bibinfo {year} {2006})\BibitemShut {NoStop}%
\bibitem [{\citenamefont {Hergert}\ \emph {et~al.}(2016)\citenamefont {Hergert}, \citenamefont {Bogner}, \citenamefont {Morris}, \citenamefont {Schwenk},\ and\ \citenamefont {Tsukiyama}}]{Ref48}%
  \BibitemOpen
  \bibfield  {author} {\bibinfo {author} {\bibfnamefont {H.}~\bibnamefont {Hergert}}, \bibinfo {author} {\bibfnamefont {S.}~\bibnamefont {Bogner}}, \bibinfo {author} {\bibfnamefont {T.}~\bibnamefont {Morris}}, \bibinfo {author} {\bibfnamefont {A.}~\bibnamefont {Schwenk}}, \ and\ \bibinfo {author} {\bibfnamefont {K.}~\bibnamefont {Tsukiyama}},\ }\href {\doibase 10.1016/j.physrep.2015.12.007} {\bibfield  {journal} {\bibinfo  {journal} {Phys. Rep.}\ }\textbf {\bibinfo {volume} {621}},\ \bibinfo {pages} {165} (\bibinfo {year} {2016})},\ \bibinfo {note} {memorial Volume in Honor of Gerald E. Brown}\BibitemShut {NoStop}%
\bibitem [{\citenamefont {Morris}\ \emph {et~al.}(2015)\citenamefont {Morris}, \citenamefont {Parzuchowski},\ and\ \citenamefont {Bogner}}]{Ref65}%
  \BibitemOpen
  \bibfield  {author} {\bibinfo {author} {\bibfnamefont {T.~D.}\ \bibnamefont {Morris}}, \bibinfo {author} {\bibfnamefont {N.~M.}\ \bibnamefont {Parzuchowski}}, \ and\ \bibinfo {author} {\bibfnamefont {S.~K.}\ \bibnamefont {Bogner}},\ }\href {\doibase 10.1103/PhysRevC.92.034331} {\bibfield  {journal} {\bibinfo  {journal} {Phys. Rev. C}\ }\textbf {\bibinfo {volume} {92}},\ \bibinfo {pages} {034331} (\bibinfo {year} {2015})}\BibitemShut {NoStop}%
\end{thebibliography}%
\end{document}